\documentclass[twoside,11pt]{article}

\RequirePackage{jmlr2e}
\RequirePackage[OT1]{fontenc}
\RequirePackage{amsmath}
\RequirePackage[colorlinks,citecolor=blue,urlcolor=blue]{hyperref}
\RequirePackage{subfigure}
\RequirePackage{bbm}

\jmlrheading{1}{publication year}{pages}{submission date}{publication date}{Mattia Ciollaro, Christopher R. Genovese, Jing Lei and Larry Wasserman}

\ShortHeadings{Mode hunting and clustering in infinite dimensions}{Ciollaro, Genovese, Lei and Wasserman}

\begin{document}

\title{The functional mean-shift algorithm for mode hunting and clustering in infinite dimensions}

\author{\name Mattia Ciollaro \email ciollaro@cmu.edu\\ 
\name Christopher R. Genovese \email genovese@stat.cmu.edu\\ 
\name Jing Lei \email jinglei@andrew.cmu.edu\\ 
\name Larry Wasserman \email larry@stat.cmu.edu\\ 
\addr Department of Statistics\\ 
Carnegie Mellon University\\ 
5000 Forbes Avenue\\
Pittsburgh (PA)\\
15213 USA}

\editor{ADD EDITOR}

\maketitle

\begin{abstract}%
We introduce the \textit{functional mean-shift algorithm}, an iterative algorithm for estimating the local modes of a surrogate density from functional data. We show that the algorithm can be used for cluster analysis of functional data. We propose a test based on the bootstrap for the significance of the estimated local modes of the surrogate density. We present two applications of our methodology. In the first application, we demonstrate how the functional mean-shift algorithm can be used to perform {\em spike sorting}, i.e. cluster neural activity curves. In the second application, we use the functional mean-shift algorithm to distinguish between original and fake signatures.\\
\end{abstract}


\section{Introduction}
\label{sec:introduction1}
\begin{figure} 
	\includegraphics[width=1\columnwidth, height=0.3\textheight]{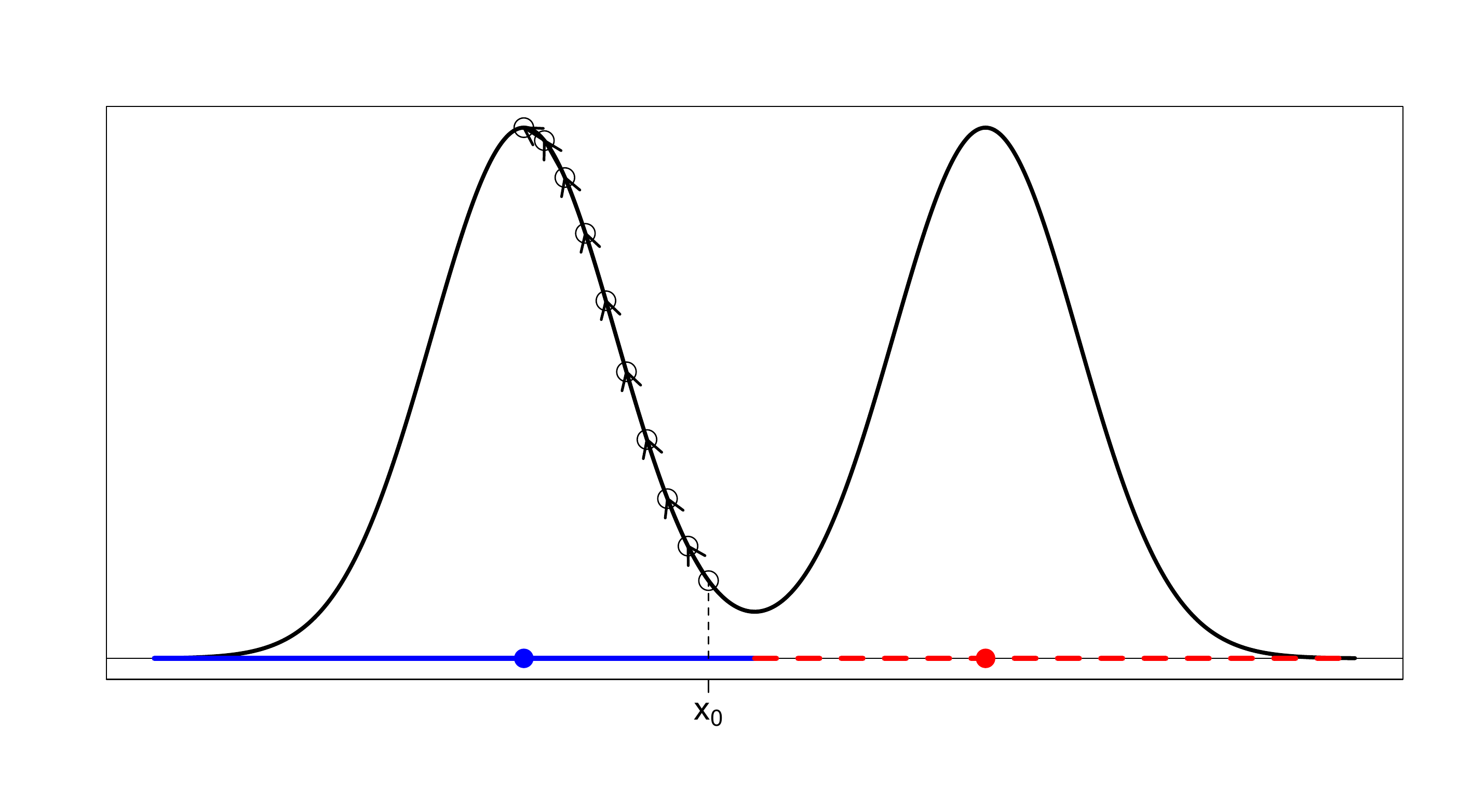}
	\caption{A probability density function representing two clusters. Each cluster is the basin of attraction of a local mode.}
	\label{fig:densityclustering}
\end{figure}
\begin{figure}
{\includegraphics[width=\columnwidth,height=0.3\textheight]{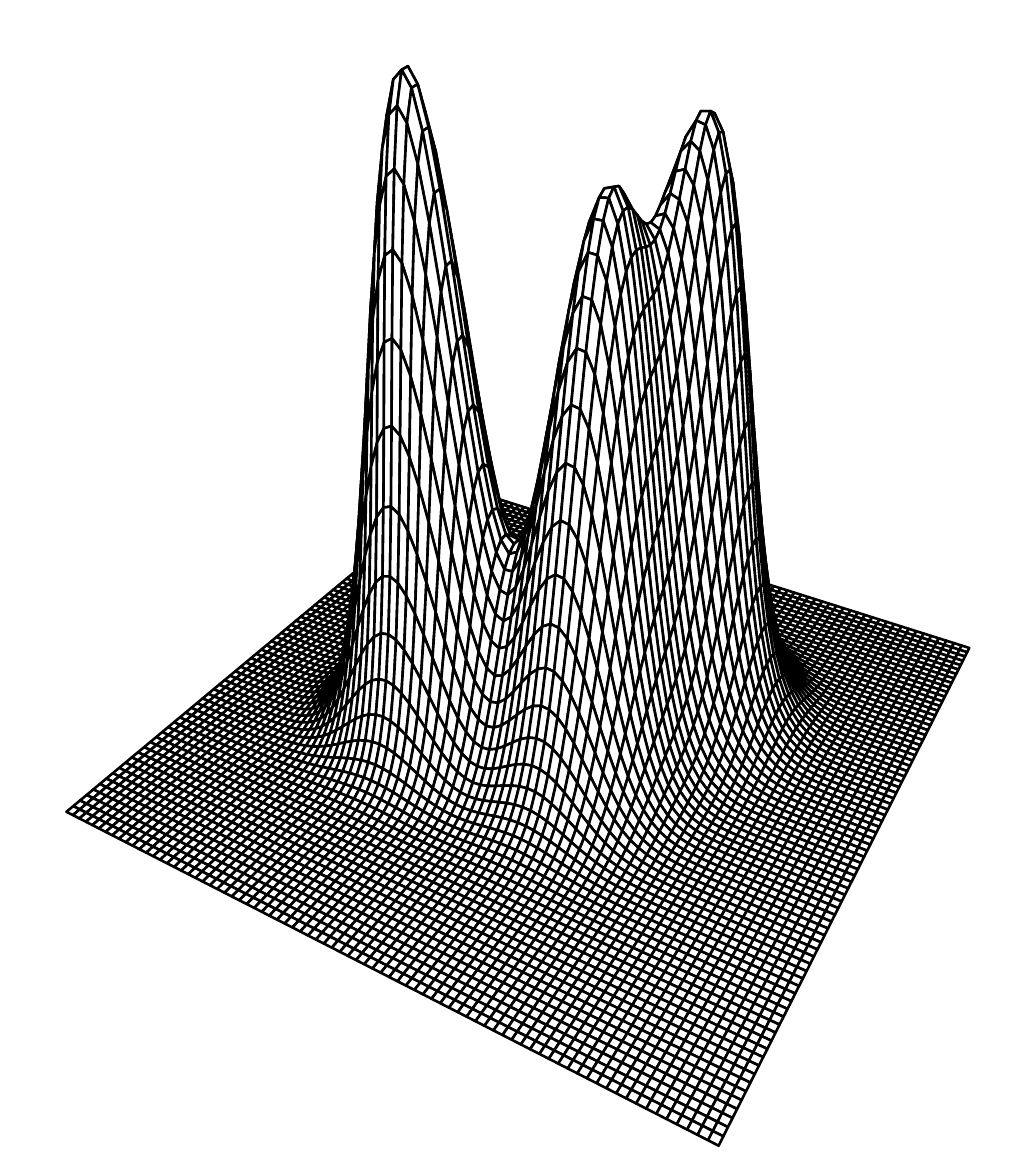}}
{\includegraphics[width=\columnwidth,height=0.5\textheight]{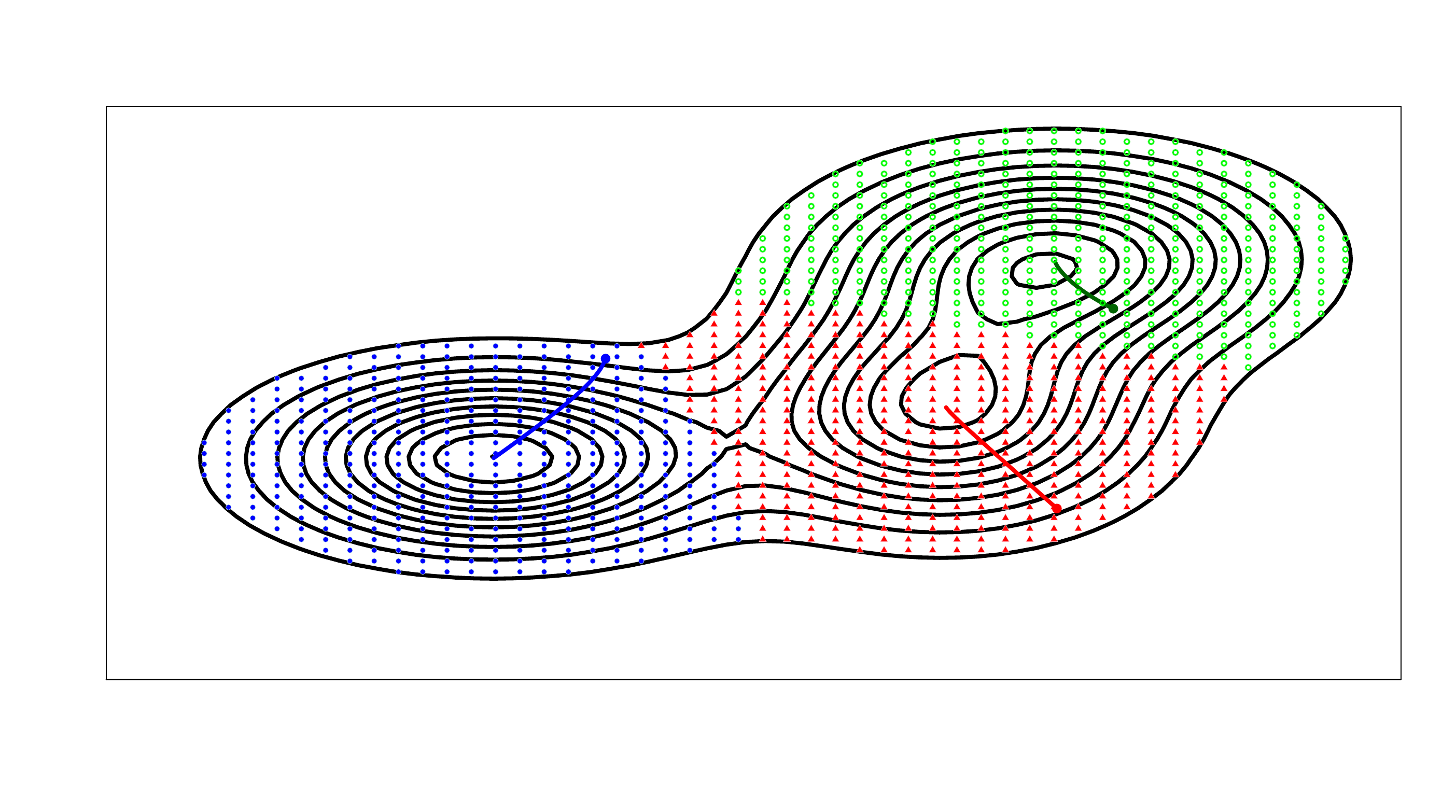}}
\caption{Modal clustering on a bivariate density with compact support. The density has three local modes and therefore three well-defined clusters. Within each basin of attraction, the gradient ascent path of a point that belongs to that basin is highligthed.}
\label{fig:2ddensityclustering}
\end{figure}

A probability density function contains information that can be used for clustering data drawn from that distribution. Figure \ref{fig:densityclustering} depicts a probability density function that is a mixture of two distinct unimodal densities. Intuitively, we can say that this density represents two clusters. To unambiguously characterize the two clusters, we can use the \textit{basin of attraction} of each local mode: if we repeatedly shift a point $x_0$ on the $x$ axis along the direction of steepest ascent, the sequence thus generated converges either to the left or to the right local mode. If the destination of $x_0$ is the left (right) local mode, then $x_0$ belongs to the basin of attraction of the left (right) local mode. The set of points in the domain of the density whose common destination is the left (right) local mode naturally form a well-defined cluster. This idea of \textit{modal clustering} easily generalizes to $q$-dimensional densities (Figure \ref{fig:2ddensityclustering}). If the density is unknown, as is usually the case in statistical inference, a set of empirical local modes and the corresponding set of empirical clusters can be obtained by performing the above procedure on an estimate of the density.

In this paper, we extend the idea of modal clustering to non-Euclidean spaces of infinite dimension. We demonstrate that:
\begin{enumerate}
\item We can meaningfully define a surrogate notion of density in an infinite dimensional space that lacks a natural dominating measure.
\item We can find the local modes of this surrogate density and the corresponding clusters.
\item We test if, given a sample of infinite dimensional data, a set of estimated local modes and their corresponding clusters are significant or only due to random fluctuation.
\end{enumerate} 
In particular,
\begin{enumerate}
\item We consider a surrogate of the notion of probability density that does not require the existence of a dominating measure. Suppose that the probability associated to the open ball of radius $\epsilon$ centered at the element $x$ of the infinite dimensional space $\mathcal{X}$ admits the asymptotic approximation
  \begin{equation*}
    P(X \in B(x,\epsilon)) = p(x)\phi(\epsilon) + o(\phi(\epsilon))
  \end{equation*}
for a functional $p$ and a function $\phi(\epsilon)$ as $\epsilon \to 0$. Then, we call the functional $p$ a surrogate density for $P$. This $p$ can be consistently estimated. 
\item We generalize the \textit{mean-shift algorithm}, a simple and fast iterative procedure that estimates the local modes of a probability density, to functional data and surrogate densities. The functional mean-shift algorithm repeatedly shift a point $x \in \mathcal{X}$ towards its closest local sample mean using the update equation
\begin{equation*}
x \leftarrow  \frac{\sum_{X \in S} k\left(\frac{d(X,x)}{h}\right) X}{\sum_{X \in S} k\left(\frac{d(X,x)}{h}\right)},   
  \end{equation*}
where $X$ is an element of the sample $\mathcal{S}=\{X_1, \dots, X_n\} \subset \mathcal{X}$, $k$ is a kernel function, $d$ is a suitable distance, and $h>0$ is a bandwidth parameter. The sequence generated by the above update equation converges to a critical point $x^*$ of a kernel density estimate of the unknown surrogate density $p$.
\item We construct a bootstrap test for the significance of the estimated local modes. Suppose that $x^*$ is a candidate local mode of the unknown surrogate density $p$.
We test whether the curvature of the surrogate density $p$ at $x^*$ is negative:
\begin{align*}
		\begin{split}
			H_0 &: \lambda_{x^*} \geq 0\\
				& \text{vs.} \\
			H_1 &: \lambda_{x^*}  < 0,
		\end{split}
\end{align*}
where $ \lambda_{x^*}=\sup_{\|y\|=1} p^{(2)}_{x^*}(y,y)$ and $p^{(2)}_{x^*}$ is the Hessian of the surrogate density $p$ at $x^*$ .
\end{enumerate} 
We illustrate our methodology and its performance with two applications on smooth curves. The first application aims at clustering a set of curves corresponding to the neural activity of a Macaque monkey performing a particular task. In the second application, we apply our methodology to distinguish between original and fake signatures.

\section{Related literature}
\label{sec:introduction2}
The problem of estimating the mode of an unknown probability density dates back at least as early as \cite{parzen1962estimation}. The work of Parzen has been extended in several directions, for instance by \cite{chernoff1964mode}, who shows the asymptotic normality of the sample mode, \cite{eddy1980}, who studies its rate of convergence, and \cite{romano1988}, who derives asymptotic minimax risk lower bounds. \cite{donoho1991geometrizing} obtain the minimax rate for estimating a mode in one dimension. \cite{vieu1996note} proposes four density mode estimators that attain the same rate of convergence of \cite{eddy1980}. \cite{klemela2005adaptive} proposes an adaptive estimator which attains the optimal rate of convergence.

The more general problem of estimating nonparametrically the local modes of an unknown probability density has its roots in the seminal work of \cite{fukunagahostetlermeanshift} on the \textit{mean-shift algorithm} (see also \citealp{silverman1981using}). The work of \cite{fukunagahostetlermeanshift}, later generalized by \cite{chengmeanshift}, also promotes the mean-shift algorithm as a tool for cluster analysis (\citealp{Silverman86}). More recently, the algorithm has been widely used in the computer science community for image segmentation tasks (\citealp{comaniciu02meanshift} and \citealp{carreira2006fast}).

The mechanics of the mean-shift algorithm are simple. An arbitrary point $x \in \mathbb{R}^q$ is repeatedly shifted towards its closest local sample mean by the iterated application of the mean-shift update equation
\begin{equation}
	\label{eq:MSupdateequation}
	x \leftarrow  \frac{\sum_{X \in S} k\left(\frac{\|X-x\|}{h}\right) X}{\sum_{X \in S} k\left(\frac{\|X-x\|}{h}\right)},
\end{equation}
where $X$ is a point belonging to the sample $\mathcal{S}=\{X_1, \dots, X_n\} \subset \mathbb{R}^q$, $k$ is a kernel function, $\| \cdot \|$ denotes the usual Euclidean norm, and $h>0$ is a bandwidth parameter. The update is performed iteratively until convergence and the sequence of $x$'s thus generated determines a polygonal line which approximates the continuous gradient ascent path joining the initial point in the sequence and the closest candidate local mode of the unknown probability density of the $X$'s. More precisely, the polygonal line generated by the repeated application of the update equation \eqref{eq:MSupdateequation} from an initial point $x \in \mathbb{R}^q$ approximates the \textit{integral curve} $\pi_x: \mathbb{R}_+ \to \mathbb{R}^q$ that solves the initial value problem
\begin{equation}
\label{eq:gradientflow}
	\begin{cases}
		\pi_x'(t) = \nabla p(\pi_x(t)) \\
		\pi_x(0) = x,
	\end{cases}
\end{equation}
where $p$ is the unknown probability density. The curve $\pi_x$ is also called a \textit{gradient flow line}. A  theoretical discussion about the connection between gradient ascent and the mean-shift algorithm, including general results on the rates of convergence of the polygonal line generated by gradient ascent to the corresponding gradient flow line, can be found in \cite{ariascastromeanshiftgradient}.

\cite{chacon2012clusters} provides a formal justification of clustering based on high density regions and local modes by means of Morse theory. The basic idea is that a \textit{cluster} can be defined as the set of points $x$ in the domain of a probability density (with non-degenerate critical points) whose gradient ascent paths $\pi_x$ culminate at the same local mode, i.e. a cluster is the \textit{basin of attraction} of a local mode (see also \citealp{li2007nonparametric}). From the theory of differential equations, it is known that a curve $\pi_x$ satisfying the initial value problem \eqref{eq:gradientflow} exists and is unique under appropriate regularity assumptions on $p$. Moreover, because distinct integral curves can only intersect at the stationary points of $p$, the equivalence class of points whose gradient flow lines culminate at the same local mode naturally forms a cluster, and the collection of these equivalence classes is a partition of the support of the density function. The definition of cluster as the basin of attraction of a local mode is particularly appealing from an inferential viewpoint, because it endows cluster analysis with a sound theoretical foundation. By embracing this definition, we have a clear dichotomy between the notion of \textit{population cluster}, which corresponds to the set of points associated to the high density region surrounding a local mode of the unknown probability density, and that of \textit{empirical cluster}, corresponding to the set of points associated to the high density region surrounding a local mode of the estimated probability density.

\cite{hartigan1975clustering} defines the population clusters of a probability distribution with density $p$ as the connected components of the upper level sets of $p$. The connected components of the set $L(\lambda)=\{x \in \mathbb{R}^q: p(x) \geq \lambda \}$ for a given $\lambda \geq 0$, are considered the population clusters of $p$. These clusters can be estimated by $\hat L(\lambda)=\{x \in \mathbb{R}^q: \hat p(x) \geq \lambda \}$, as soon as an estimate $\hat p$ of $p$ is available, or otherwise by means of some other estimate of $L(\lambda)$  (see, among others, \citealp{cuevas1997plug}; \citealp{stuetzle2003estimating}; \citealp{azzalini2007clustering}; \citealp{stuetzle2010generalized}; \citealp{rinaldo2010generalized}). However, the fact that this definition of population cluster depends on the resolution parameter $\lambda$ is a severe limitation and may represent a drawback in certain applications (see, for instance, \citealp{rinaldo2012stability}). Furthermore, the computation of plug-in estimates of $L(\lambda)$ and their connected components poses some difficulties and often requires intricate algorithms. Finally, a completely different approach to mode-based clustering based on persistent homology is considered by \cite{chazal2013persistence}.

Although the problem of estimating the local modes of a probability density and using them for cluster analysis has been extensively studied in the Euclidean case, and the mean-shift algorithm is widely used to perform this task, the same challenge has not received the same attention with \textit{functional data}. Functional Data Analysis (\citealp{bosq2000linear}; \citealp{ramsay2006functional}; \citealp{ferraty2006nonparametric}; \citealp{horvath2012inference}) is a modern branch of statistics which deals with data that are intrinsically infinite dimensional such as functions, curves or surfaces. In the last two decades, several statistical methods (both parametric and nonparametric) that are fully developed for Euclidean data have been extended to the setting of random variables with infinite dimensional realizations. The theory of mode estimation for functional data is recent (see, for instance, \citealp{gasser1998nonparametric}; \citealp{dabo2004estimation} or \citealp{ferraty2006nonparametric} and references therein), but the estimation of \emph{local} modes for functional data has received less attention, and a principled extension of the mean-shift algorithm that is adapted to this type of data has not been devised. Similarly, the literature on clustering methods for functional data is not as vast as that for Euclidean data (the reader can refer to \citealp{jacques2013functional} for a recent review). In particular, density-based clustering has received scant attention in the setting of functional data, mainly because of the difficulty related to defining a probability density in infinite dimensional spaces. In this paper, we address these challenges by providing a statistical framework that allows us to define and interpret a mean-shift algorithm for functional data, which we call the \textit{functional mean-shift algorithm}.


\section{The mean-shift algorithm for Euclidean data}
\label{sec:meanshift}

The original mean-shift algorithm was proposed by \cite{fukunagahostetlermeanshift}. The algorithm consists of repeatedly shifting a finite set of data points in the $q$ dimensional Euclidean space towards their local sample means. More precisely, let $\mathcal{S}=\{X_1, \dots, X_n\} \subset \mathbb{R}^q$ be a sample of i.i.d. random vectors with distribution $P$ that has density $p$ with respect to the Lebesgue measure. Then, \cite{fukunagahostetlermeanshift} define the mean-shift operator $m(\cdot)$ as
\begin{equation}
\label{eq:originalmeanshiftoperator}
	m(x) = \frac{\sum_{X \in S} K_h (X-x) X}{\sum_{X \in S} K_h(X-x)} - x,
\end{equation}
where $K$ is a kernel function and $K_h(\cdot)=k(\| \cdot \|/h)$. For instance, $K$ may correspond to the boxcar kernel
\begin{equation*} 
	K(x) 	= k(\|x\|) =
		\begin{cases}
			1	&\text{if } \|x\| \leq 1 \\
			0	&\text{if } \|x\| > 1
		\end{cases},
\end{equation*}
or the Gaussian kernel
\begin{equation*} 
	\label{eq:gaussiankernel}
	K(x) 	= k(\|x\|) \propto e^{-\frac{\|x\|^2}{2}}.
\end{equation*}
Here, $\|\cdot\|$ indicates the standard Euclidean norm, $h>0$ is a bandwidth parameter and
\begin{equation*}
	x + m(x) = \frac{\sum_{X \in S} K_h(X-x) X}{\sum_{X \in S} K_h(X-x)}
\end{equation*}
is a local sample mean. For a general kernel function $K(x)=k(\|x\|)$, the associated function $k$ is referred to as the {\em profile} of $K$.

The original mean-shift algorithm consists of repeatedly updating all the data points $X \in \mathcal{S}$ simultaneously according to the mean-shift update
\begin{equation}
	\label{eq:originalmeanshiftupdate}
	X \leftarrow X + m(X)
\end{equation}
until convergence. Because the data points $X \in \mathcal{S}$ tend to converge to a finite number of points $\tilde{\mathcal{M}}=\{\tilde \mu_1, \dots, \tilde \mu_r\} \subset \mathbb{R}^q$, the algorithm can be used to perform clustering. In particular, the $j$-th cluster of data points is defined as the subset of $X$'s in $\mathcal{S}$ all converging to $\tilde \mu_j$ after the repeated iteration of the mean-shift update.

\cite{chengmeanshift} generalizes the original mean-shift algorithm proposed by \cite{fukunagahostetlermeanshift} and shows that the mean-shift algorithm includes $k$-means clustering as a limit case. He considers general isotropic kernels $K$ and allows for the presence of positive weights $\omega$ in the local averaging of the sample points, so that the mean-shift operator of equation \eqref{eq:originalmeanshiftoperator} now reads
\begin{equation}
	\label{eq:meanshiftoperatorweights}
	m(x) = \frac{\sum_{X \in S} \omega(X) K_h(X-x) X}{\sum_{X \in S} \omega(X) K_h(X-x)} - x.
\end{equation}
Most importantly, \cite{chengmeanshift} allows the mean-shift operator $m$ to act on any arbitrary subset of points in $\mathbb{R}^q$ while keeping the sample $\mathcal{S}$ fixed, as opposed to the original \cite{fukunagahostetlermeanshift} version which applies equation \eqref{eq:originalmeanshiftupdate} simultaneously to all the observed data points $X \in \mathcal{S}$, in fact updating the entire sample $\mathcal{S} \leftarrow \mathcal{S} + m(\mathcal{S})$ and thus generating a `blurring' sequence of multi-sets. Today, Cheng's version of the algorithm is called the mean-shift algorithm, and the original algorithm of Fukunaga and Hostetler is instead referred to as the blurring mean-shift algorithm.

When using Cheng's mean-shift algorithm, one can imagine following the path generated by the update equation
\begin{equation}
	\label{eq:meanshiftupdate}
	x \leftarrow x + m(x) = \frac{\sum_{X \in S} \omega(X) K_h(X-x) X}{\sum_{X \in S} \omega(X) K_h(X-x)}
\end{equation}
from any arbitrary initial $x \in \mathbb{R}^q$ while keeping the data points $X \in \mathcal{S}$ fixed. This operation corresponds to performing gradient ascent on the kernel density estimate
\begin{equation*}
	\tilde p(x) = \frac{1}{nh^q} \sum_{X \in \mathcal{S}} \omega(X) G_h(X-x)
\end{equation*}
of the true density $p$ of the $X$'s based on a second kernel $G$ (often called the {\em shadow} of $K$). It is easy to verify that the mean-shift update of equation \eqref{eq:meanshiftupdate} corresponds to an update of the type
\begin{equation*}
	x \leftarrow x + s(x)  \nabla \tilde p(x),
\end{equation*}
where $s(x)$ is a step size parameter depending on the current position $x$, and $\nabla \tilde p(x)$ corresponds to the gradient of $\tilde p $ at $x$. Equivalently, one can rewrite the update above as
\begin{equation}
	\label{eq:steepestascentupdate_normalized}
		x \leftarrow x + s(x) \| \nabla \tilde p(x) \| a^*(x) = x + \bar{s}(x) a^*(x),
\end{equation}
where the adaptive step size is
\begin{equation}
	\label{eq:euclideanstepsize}
	\bar{s}(x) \propto \| \nabla \tilde p(x) \| \frac{h^2}{\hat p(x)},
\end{equation}
$a^*$ is the unitary norm vector in the direction of the gradient $\nabla \tilde p(x)$ and
\begin{equation*}
	\hat p(x) = \frac{1}{nh^q} \sum_{X \in \mathcal{S}} \omega(X) K_h(X-x)
\end{equation*}
is  a kernel density estimate of the unknown density $p$ using the kernel $K$. The unit norm vector $a^*$ gives the direction of steepest ascent at $x$. The important feature of equation \eqref{eq:euclideanstepsize} is the adaptive nature of the step size: if the current position $x$ corresponds to a low density position (i.e. $x$ is far from a local mode), then $\hat p(x)$ is small and the step in the direction of the gradient ascent is large; conversely, if the current position $x$ corresponds to a high density position (i.e. $x$ already is close to a local mode), then $\hat p(x)$ is large and the step in the direction of gradient ascent is small. Thus, thanks to the adaptivity of the step size, low density regions are only visited for a small number of iterations while, at the same time, the algorithm does not overshoot in high density regions.

The repeated application of the mean-shift update starting from an initial arbitrary point $x \in \mathbb{R}^q$ generates a trajectory that eventually converges to a (candidate) local mode of the estimated density $\tilde p$. This trajectory, which is a polygonal line, can be viewed as an estimate of the unknown gradient ascent path (or \textit{gradient flow line}) $\pi_x: \mathbb{R}_+ \to \mathbb{R}^q$ that solves the initial value problem
\[
\begin{cases}
	\pi_x'(t) = \nabla p(\pi_x(t)) \\
	\pi_x(0) = x.
\end{cases}
\]
This feature of the algorithm allows us to determine a partition of $\mathbb{R}^q$ in which each set of the partition corresponds to the basin of attraction of a local mode of the estimated density $\tilde p$, i.e. the equivalence class of all the points $x \in \mathbb{R}^q$ whose mean-shift trajectories culminate at the same local mode $\tilde \mu$ of $\tilde p$ (which is an estimate of the equivalence class of points $x \in \mathbb{R}^q$ whose unknown gradient flow lines of $p$ converge to the same unknown local mode $\mu$ of $p$). Similarly, a cluster of data points in the sample $\mathcal{S}$ can be defined as a subset of $X$'s in $\mathcal{S}$ whose mean-shift trajectories culminate at the same local mode of the estimated density $\tilde p$.

If the true density $p$ was known, both the location and the number of the local modes $\mathcal{M}=\{\mu_1, \dots, \mu_R\}$ of $p$ would be known. Therefore, if one accepts the above definition of cluster as basin of attraction of a local mode of the density, knowing $p$ implies knowing the true {\em population} clustering of $\mathbb{R}^q$ based on $p$. Because the true density $p$ is generally unknown, the set $\mathcal{M}$ (and therefore the true population clusters) are instead estimated by performing gradient ascent on an estimate $\tilde p$ of the true density $p$. The estimated local modes $\tilde{\mathcal{M}}=\{ \tilde \mu_1, \dots, \tilde \mu_r \}$ and their corresponding basins of attraction based on $\tilde p$ thus provide a natural estimate of the unknown true population clustering. Recent results about the convergence of empirical clusters to their corresponding population clusters have been obtained by \cite{chacon2012populationbackground}.
\section{A statistical framework for the functional mean-shift algorithm}
\label{sec:surrogatedensities}
In the remainder of the paper, we assume that the data $\mathcal{S}=\{X_1,\dots,X_n\}$ consist of $n$ i.i.d. random variables defined on a suitable abstract probability space $(\Omega, \mathcal{A}, P)$ and we further assume that each random variable $X \in \mathcal{S}$ takes values in a (potentially) infinite dimensional measurable Hilbert space $(\mathcal{X},\langle \cdot,\cdot \rangle)$. The inner product induces a norm $\|\cdot\|$ on $\mathcal{X}$, defined for any element $x \in \mathcal{X}$ as $\|x\|=\sqrt{\langle x, x \rangle}$. The norm $\|\cdot\|$, in turn, generates a distance $d$, defined between any pair $x,y$ of elements of $\mathcal{X}$ as $d(x,y)=\|x-y\|$.

Often it can be difficult to define a density for $P$ if $\mathcal{X}$ lacks a natural dominating measure. Whenever this is the case, we replace the notion of density with that of a \textit{surrogate density} as in \cite{gasser1998nonparametric} and \cite{ferratysurrogatedensity}. Specifically, let $B(x,\epsilon) \subseteq \mathcal{X}$ indicate the open ball of radius $\epsilon$ centered at $x$ in the topology induced by $d$ and assume that there exist a function $\phi$ and a well-behaved functional $p$ such that for any $x \in \mathcal{X}$
\begin{equation}
	\label{eq:smallballdecomposition}
	\varphi_x(\epsilon) = P(X \in B(x,\epsilon)) \sim p(x)\phi(\epsilon),
\end{equation}
meaning that the so-called {\em small ball probability function} $\varphi_x(\epsilon)$ satisfies $\varphi_x(\epsilon)=p(x)\phi(\epsilon) + o(\phi(\epsilon))$ as $\epsilon \to 0$. Then, we consider the functional $p: \mathcal{X} \to \mathbb{R}_+$ satisfying the identifiability condition $E(p(X))=1$ a surrogate probability density of $P$ on $\mathcal{X}$, and the function $\phi$ (often referred to as the \textit{concentration function}) can be considered a volume parameter. \cite{estimating2006ferraty} and \cite{ferratysurrogatedensity} provide some examples of stochastic processes whose small ball probabilities fulfill the decomposition of equation \eqref{eq:smallballdecomposition}. Notice that in the simpler setting in which $X$ is a continuous real valued random variable with cumulative distribution function $F$, one has $\varphi_x(\epsilon)=P(X \in B(x,\epsilon))=F(x+\epsilon)-F(x-\epsilon)=p(x) 2\epsilon + o(\epsilon)$ where $p$ here corresponds to the usual Radon-Nikodym derivative of $P$ with respect to the Lebesgue measure and $\phi(\epsilon)=2\epsilon$ is the Lebesgue measure of $B(x,\epsilon)$. However, for our purposes, we do not assume the existence of a dominating measure for $P$.

\cite{ferratysurrogatedensity} propose to estimate the population surrogate density $p$ by means of
\begin{equation}
	\label{eq:surrogateestimator}
		\hat p(x)=\frac{\frac{1}{n} \sum_{i=1}^n K_h \left(X_i,x \right)}{\frac{1}{n(n-1)} \sum_{i=1}^n\sum_{j \neq i}K_h \left( X_i, X_j \right)} = w_K(\mathcal{S}) \sum_{X \in \mathcal{S}} K_h \left( X,x\right),
\end{equation}
which is well-defined for $h \geq \min_{i,j} d(X_i,X_j)$. The kernel $K$ is defined in terms of its profile $k$ as
\begin{equation}
	\label{eq:kernel}
	K(x,y) = k(d(x,y)),
\end{equation}
and the profile $k$ is a function of the distance $d$ between two points $x$ and $y$. The estimator is consistent under mild assumptions \cite[assumptions H1--H4]{ferratysurrogatedensity}. We henceforth assume that $k \geq 0$, $k$ has compact support $[0,1]$, $\int_\mathbb{R} k(t)\,dt=1$ (although, as pointed out by \citealp{chengmeanshift}, the unit integral condition is superfluous for the mean-shift algorithm), there exist constants $-\infty < C_1 < C_2 \leq 0$ such that $C_2 \leq k' \leq C_1$ in $(0,1)$ and $k''$ exists in $(0,1)$.

It is worth mentioning that other surrogate densities have been proposed for random variables valued in infinite dimensional spaces. For instance, the reader may refer to \cite{delaigle2010} who define a surrogate probability density for function-valued random variables on the basis of the eigendecomposition obtained by means of principal component analysis.
\begin{remark}
Assuming that the factorization of the small ball probability of equation \eqref{eq:smallballdecomposition} holds is not strictly necessary in practice, although it helps to establish a clearer connection between the standard mean-shift algorithm and the functional mean-shift algorithm in the rest of our discussion. When the factorization does not hold, one can work with the numerator of the estimator of equation \eqref{eq:surrogateestimator},
\[
	\frac{1}{n} \sum_{X \in \mathcal{S}} K_h \left(X,x \right)
\]
and view it as an estimator of its own expected value, $\mathbb{E}_P\left( K_h\left(X,x\right)\right)$. In turn, $\mathbb{E}_P\left( K_h\left(X,x\right)\right)$ can be interpreted as a smooth population functional which (for some $h>0$) is informative about local features of $P$.
\end{remark}
\begin{remark}
The choice of the distance in equation \eqref{eq:kernel} should not be perceived as a complication, but rather as a useful element of flexibility in applied work. While on the one hand more traditional methods implicitly impose to the user the choice of the space in which to embed the data (e.g. functional PCA forces the user towards $L_2$ spaces and $L_2$ distances), the functional mean-shift algorithm allows the user to incorporate prior knowledge of the data generating process by choosing in which space to embed the data. This is implicitly done by tuning the distance function. In many practical problems, this flexibility can save a lot of work in pre-processing the data. For example, suppose that we observe a sample of curves that exhibit the same basic pattern except for a completely random and uninformative vertical shift, i.e. $X_i(t) = X_j(t) + \delta_{ij}$ where $\delta_{ij} \in \mathbb{R}$. In this case, a (semi-)distance based on the first derivatives such as
\[
	d(x,y)=\| x'-y'\|_{L_2}
\]
avoids the need to perform a vertical alignment of the curves before further analysis. In fact, in this case, $d(X_i(t),X_j(t))=0$. Similar considerations apply for systematic differences in the higher order derivatives among the curves (see \citealp{ferraty2006nonparametric}).
\end{remark}

\section{The functional mean-shift algorithm}
\label{sec:meanshiftfunctionspaces}
In this section, we show that applying the mean-shift algorithm to the elements of an Hilbert space using a kernel $K$ is equivalent to applying an adaptive gradient ascent algorithm based on the Gate\^aux derivative of the estimated surrogate density 
\begin{equation*}
\tilde p(x) = w_G(\mathcal{S}) \sum_{X \in \mathcal{S}} G_h \left( X,x\right)
\end{equation*} 
based on a second kernel, $G$. The kernel $G$, which is often referred to as the \textit{shadow} of $K$, has a profile $g$ which is related to the profile $k$ of $K$ in a particular way. We further discuss the notion of shadow of a kernel and the link between the profile $k$ and $g$ later in this section. We show that the gradient ascent direction can be characterized as the element of $\mathcal{X}$ that has unit norm and maximizes the Gate\^aux derivative of $\tilde p$. 

Before proceeding, let us briefly recall the definition of Gate\^aux differential and derivative.
\begin{definition}[Gate\^aux differentiability, \citealp{ambrosetti1995primer}]
Let $\mathcal{B}$ be a a Banach space and let $U$ be an open subset of $\mathcal{B}$. The map $F:U \mapsto V$ is Gate\^aux differentiable at $u \in U$ if there exists a continuous linear map $F_u$ from $\mathcal{B}$ to $V$, called the Gate\^aux differential of $F$ at $u$, such that for all $w \in U$
\begin{equation*}
	\lim_{\alpha \to 0} \frac{F(u+\alpha w)-F(u)}{\alpha} = F_u(w),
\end{equation*}
where $\alpha \in \mathbb{R}$. If $F$ is Gate\^aux differentiable at all $u \in U$, then $F$ is said to be Gate\^aux differentiable in $U$. The map $F'_G: u \mapsto F_u$, which associates to each $u \in U$ the continuous linear operator $F_u$, is called the Gate\^aux derivative of $F$.
\end{definition}
The Gate\^aux derivative of the estimated surrogate density $\tilde p$ is obtained in the following Lemma.
\begin{lemma}[Gate\^aux differential of the estimated surrogate density]
\label{lemma:gateaux}
Let $\tilde p_x(y)$ denote the Gate\^aux differential of $\tilde p$ at $x \in \mathcal{X}$ evaluated at $y \in \mathcal{X}$. Let $h(X)$ be an adaptive bandwidth possibly varying across the data points $X \in \mathcal{S}$ which possibly depends on the entire sample $\mathcal{S}$, but not on $x$. Then,
\begin{equation}
	\label{eq:gateaux}
	\tilde p_x(y) = \left \langle y, Cw_G(\mathcal{S}) \sum_{X \in \mathcal{S}} \frac{1}{h^2(X)} K_h \left( X,x\right) \left(X-x\right) \right \rangle,
\end{equation}
where the profiles $g$ and $k$ of the kernel $G$ used in $\tilde p$ and the kernel $K$ satisfy $k(t)=-g'(t)/(Ct)$ for a constant $C>0$.
\end{lemma}
\begin{remark}
If the profile $k$ of $K$ is not continuous at 1 (as it is the case for the truncated Gaussian profile for instance), the Gate\^aux derivative of $\tilde p$ is still almost surely continuous at any fixed $x \in \mathcal{X}$ as long as $P\left( \bigcup_{X \in \mathcal{S}} \left\{ d(X,x)=h(X) \right\}\right)=0$.
\end{remark}
In the second last line of the proof of the above Lemma (see Appendix), we set
\begin{equation}
	\label{eq:kernelchange}
	k(t)=-\frac{g'(t)}{Ct}
\end{equation}
with
\begin{equation*}
	C=\int_{\mathbb{R}} -\frac{g'(t)}{t}\,dt > 0.
\end{equation*}
One can verify that equation \eqref{eq:kernelchange} (which is a differential version of equation (15) of \citealp{chengmeanshift}) implies, among others, the correspondences of Table \ref{tab:kernelcorrespondences}. For our purposes, we give the following definition of the shadow of a kernel.
\begin{definition}[Shadow of a kernel]
If the profile $k$ of the kernel $K$ used in the mean-shift algorithm and the profile $g$ of the kernel $G$ used in the corresponding gradient ascent algorithm satisfy equation \eqref{eq:kernelchange}, then $G$ is said to be the shadow of $K$.
\end{definition}
As a matter of fact, Gaussian kernels are the only kernels that coincide with their own shadows; see Table \ref{tab:kernelcorrespondences} and \cite[Theorem 2]{chengmeanshift}.

When the Gate\^aux derivative of a functional $F: \mathcal{X} \to \mathbb{R}$ at a point $x \in \mathcal{X}$ can be expressed as an inner product of the form $\langle y, \nabla \tilde p_x \rangle$ for any $y \in \mathcal{X}$, then the element $\nabla \tilde p_x \in \mathcal{X}$ is the {\em functional gradient of $\tilde p$ at $x$}. Thus, the function
\begin{equation}
	\label{eq:steepestascentdirection}
	\nabla \tilde p_x = Cw_G(\mathcal{S}) \sum_{X \in \mathcal{S}} \frac{1}{h^2(X)} K_h \left( X,x\right) (X-x)
\end{equation}
of equation \eqref{eq:gateaux} is the functional gradient of $\tilde p$ at $x$.
\begin{table*}
\caption{Some correspondences between the profile $k$ of the kernel $K$ and the profile $g$ of its shadow $G$ implied by equation \eqref{eq:kernelchange}.}
\label{tab:kernelcorrespondences}
\begin{center}
	\begin{tabular}{|c|c|}
	\hline
		$G$																&	$K$ \\
	\hline
		$g(t) \propto (1-t^2)\mathbbm{1}_{[0,1]}(t)$ (Epanechnikov)					&	$k(t) \propto \mathbbm{1}_{[0,1]}(t)$ (uniform)	\\
		$g(t) \propto (1-t^2)^2\mathbbm{1}_{[0,1]}(t)$ (biweight)						&	$k(t) \propto (1-t^2)\mathbbm{1}_{[0,1]}(t)$ (Epanechnikov)	\\
		$g(t) \propto (1-t^2)^3\mathbbm{1}_{[0,1]}(t)$ (triweight)						&	$k(t) \propto (1-t^2)^2\mathbbm{1}_{[0,1]}(t)$ (biweight)	\\
		$g(t) \propto \cos \left( \frac{\pi}{2} t \right)\mathbbm{1}_{[0,1]}(t)$ (cosine)			&	$k(t) \propto \frac{\sin \left( \frac{\pi}{2}t \right)}{t}\mathbbm{1}_{[0,1]}(t)$ (sinc) \\
		$g(t) \propto e^{-\frac{t^2}{2}}\mathbbm{1}_{[0,1]}(t)$ (Gaussian)				&	$k(t) \propto e^{-\frac{t^2}{2}}\mathbbm{1}_{[0,1]}(t)$ (Gaussian)	\\
	\hline
	\end{tabular}
\end{center}
\end{table*}

We can rewrite equation \eqref{eq:steepestascentdirection} as
\begin{equation}
	\label{eq:steepestascentdirection2}
	\nabla \tilde p_x = Cw_G(\mathcal{S}) m(x) \bar p(x),
\end{equation}
where $\bar p$ is a weighted and unnormalized estimate of the surrogate density $p$ based on $K$,
\begin{equation*}
\bar p(x) = \sum_{X \in \mathcal{S}} \frac{1}{h^2(X)} K_h(X,x),
\end{equation*}
and $m$ is the functional mean-shift operator
\begin{equation}
	\label{eq:functionalmeanshiftoperator}
	m(x) = \frac{\sum_{X \in \mathcal{S}} \frac{1}{h^2(X)} K_h(X,x)X}{\sum_{X \in \mathcal{S}} \frac{1}{h^2(X)} K_h(X,x)}-x.
\end{equation}
Notice that equation \eqref{eq:functionalmeanshiftoperator} is a functional analog of equation \eqref{eq:meanshiftoperatorweights} with $\omega(X)=1/h^2(X)$. Let $a^*(x)$ denote the element with unit norm in the direction of $\nabla \tilde p_x$ from $x$, i.e. 
\begin{equation*}
	a^*(x)=\frac{\nabla \tilde p_x}{\| \nabla \tilde p_x \|}.
\end{equation*}
Then, from \eqref{eq:steepestascentdirection2} it follows that
\begin{equation}
	\label{eq:meanshiftissteepestascent}
	m(x) = \frac{\sum_{X \in \mathcal{S}} \frac{1}{h^2(X)} K_h(X,x)X}{\sum_{X \in \mathcal{S}} \frac{1}{h^2(X)} K_h(X,x)}-x = \frac{ \| \nabla \tilde p_x \|}{C w_G(\mathcal{S}) \bar p(x)} a^*(x),
\end{equation}
hence the functional mean-shift update
\begin{equation}
	\label{eq:functionalmeanshiftupdate}
	x \leftarrow x +  m(x) = \frac{\sum_{X \in \mathcal{S}} \frac{1}{h^2(X)} K_h(X,x)X}{\sum_{X \in \mathcal{S}} \frac{1}{h^2(X)} K_h(X,x)}
\end{equation}
corresponds to a gradient ascent update $x \leftarrow x + s(x) a^*(x)$ that is conceptually identical to equation \eqref{eq:steepestascentupdate_normalized}, with a step size
\begin{equation}
	\label{eq:functionalstepsize}
	s(x) =  \frac{ \| \nabla \tilde p_x \| }{C w_G(\mathcal{S}) \bar p(x)}
\end{equation}
that is conceptually identical to \eqref{eq:euclideanstepsize}. Notice further that if $x^* \in \mathcal{X}$ is a fixed point of the functional mean-shift update of equation \eqref{eq:functionalmeanshiftupdate} (i.e. $m(x^*)=0$ in equation \eqref{eq:meanshiftissteepestascent}), then it is easily seen that $\nabla \tilde p_{x^*}=0$.

It is now clear that, because the functional mean-shift algorithm is a gradient ascent algorithm in an infinite-dimensional space, the trajectory generated by the repeated application of the functional mean-shift operator of equation \eqref{eq:functionalmeanshiftupdate} can be thought of as an estimate of the gradient flow line associated to the initial value problem
\[
\begin{cases}
	\pi_x'(t) = \nabla p_{\pi_x(t)} \\
	\pi_x(0) = x,
\end{cases}
\]
whenever such flow exists. It is known from the theory of differential equations with Lipschitz coefficients that, in a generic Hilbert space $\mathcal{X}$, the initial value problem above has exactly one solution for any starting point $x \in \mathcal{X}$ if functional gradient of $p$ is a Lipschitz map, i.e. if there exists $L>0$ such that for any $x,y \in \mathcal{X}$ we have $\|\nabla p_x - \nabla p_y \| \leq L \|x-y\|$. It is easy to check that the functional gradient of the estimated surrogate density is a Lipschitz map, therefore the functional gradient flow starting from an arbitrary point $x \in \mathcal{X}$ exists and it is unique on the estimated surrogate density. Furthermore, because any finite sample only spans a finite dimensional subspace of $\mathcal{X}$, the gradient flows on the estimated surrogate density also converge. The Lipschitz condition, however, is not enough to guarantee the convergence of the gradient flows as $t \to \infty$ in general, and in particular it does not guarantee the convergence of the gradient flows of a general population surrogate density $p$. Convergence, indeed, is a delicate question (especially in infinite dimensional spaces) and usually requires strong regularity conditions on $p$. We do not address this question in this paper, nor do we address the question of the convergence of the sequences generated by the mean-shift algorithm (which is a topic of active research even in finite dimensions; see, among others,  \citealp{LiNoteOnConvergence} and \citealp{AliyariGhassabehOnConvergence}). Intuitively, however, a sequence of elements of $\mathcal{X}$ obtained by means of the functional mean-shift algorithm approximates the corresponding gradient flow on the estimated surrogate density $\tilde p$, and can be thought of as an estimate of the associated gradient flow on the unknown population surrogate density $p$ (whenever such flow exists).
\begin{remark}
If the bandwidth is fixed rather than adaptive, i.e. $h(X)=h>0$ for all $X \in \mathcal{S}$, then the step size of equation \eqref{eq:functionalstepsize} is
\begin{equation*}
	s(x) =  \frac{ \| \nabla \tilde p_x \| }{C w_G(\mathcal{S})} \frac{h^2}{ \bar{ \bar p}(x)} \propto \| \nabla \tilde p_x \| \frac{h^2}{ \bar{\bar p}(x)},
\end{equation*}
where now the fixed bandwidth is not incorporated in the unnormalized surrogate density estimate
\begin{equation*}
	\bar{ \bar p}(x) = \sum_{X \in \mathcal{S}} K_h(X,x),
\end{equation*}
and the analogy with equation \eqref{eq:euclideanstepsize} is even more evident.
\end{remark}
\begin{remark}
The first order condition
\begin{equation*}
	Cw_G(\mathcal{S})\sum_{X \in \mathcal{S}} \frac{1}{h^2(X)} K_h \left( X,x\right) (X-x) = 0
\end{equation*}
is a necessary optimality condition that is satisfied by all critical points of the surrogate density estimate $\tilde p$, and by the uninteresting trivial roots in the set $\mathcal{X} \setminus \bigcup_{X \in \mathcal{S}} \bar B(X, h(X))$, where $\bar B(X,h(X))$ denotes the closure of the open ball. This condition is not sufficient for local maxima and thus every $x \in \bigcup_{X \in \mathcal{S}} \bar B(X, h(X))$ satisfying $\nabla \tilde p_x = 0$ should be considered a \textit{candidate} local mode of the estimated surrogate density $\tilde p$.
To verify that a solution $x^* \in \bigcup_{X \in \mathcal{S}} \bar B(X, h(X))$ of the functional equation $\nabla \tilde p_x = 0$ is in fact a local mode and rule out unstable maxima (such as saddle points or plateaus), one can apply a suitable perturbation $\delta$ to $x^*$ and then reapply the functional mean-shift algorithm to $x^* + \delta$. If the sequence of functions generated by the functional mean-shift algorithm starting from $x^* + \delta$ still converges to $x^*$, then $x^*$ can be regarded as a local mode of the estimated surrogate density.

In the Euclidean setting, \cite{genovese2013nonparametric} develop a procedure to test the hypothesis \textit{$H_0: x^*$ is not a mode of the population density $p$} versus \textit{$H_1:x^*$ is a mode of the population density $p$}. We provide a similar test for the functional case in the next section.
\end{remark}
\begin{remark}
\label{remark:conditionsonprofiles}
Equation \eqref{eq:kernelchange} imposes some restrictions on the choice of the profiles $k$ and $g$. In fact, once we fix the profile $g$ in the gradient ascent scheme, the profile $k$ of the corresponding mean-shift algorithm must satisfy the differential inequality
\begin{equation}
	\label{eq:differentialinequality}
	k'(t) = \frac{tg''(t)-g'(t)}{t^2} \leq 0
\end{equation}
for $t \in (0,1)$ and the condition
\begin{equation}
	\label{eq:limitconditionat0}
	\lim_{t \to 0^+} k(t) = \lim_{t \to 0^+} - \frac{g'(t)}{t} = c
\end{equation}
with $0 < c < \infty$. In particular, \eqref{eq:differentialinequality} implies that $g$ must satisfy
\begin{equation*}
	g(t) \leq c_1 + c_2t^2
\end{equation*}
for some $c_1>0$ (to ensure that $g$ is positive at $t=0$ and $c_2<0$ (to ensure that $g$ is decreasing for $t \geq 0$), while \eqref{eq:limitconditionat0} requires that $g$ is locally quadratic
in a (right) neighborhood of $t=0$. All the profiles presented in Table \ref{tab:kernelcorrespondences} satisfy these two conditions.
\end{remark}
\begin{remark}
\label{remark:outliers}
Another useful property of the mean-shift algorithm is that it automatically detects outliers in the sample $\mathcal{S}$. If an element $X^* \in \mathcal{S}$ is separated from the other elements $X \in \mathcal{S}$ with respect to the norm $\| \cdot \|$ of the Hilbert space in which the data are embedded, then the algorithm generates the atomic cluster $\{X^*\}$. To illustrate, suppose that we are able to choose the bandwidths $h(X)$ according to a `good' procedure. Then, if
\[
	X^* \notin \bigcup_{X \in \mathcal{S}} \bar B(X, h(X)),
\]
we have $\nabla \tilde p_{X^*}=Cw_G(\mathcal{S}) \sum_{X \in \mathcal{S}} \frac{1}{h^2(X)} K_h \left( X,X^*\right) (X-X^*)=0$ as long as $K$ has a profile that is supported on $[0,1]$. Atomic clusters produced by the functional mean-shift algorithm under a `good' choice of the bandwidth should therefore be regarded as potential outliers.
\end{remark}

We conclude this section by noting that, in analogy with the Euclidean case, the blurring version of the functional mean-shift algorithm is easily obtained by iterating
\begin{equation*}
	X \leftarrow X + m(X)
\end{equation*}
on all $X \in \mathcal{S}$ simultaneously, with $m$ as in equation \eqref{eq:functionalmeanshiftoperator}.

\section{Connection with clustering using fPCA and $k$-means}
\label{sec:comparisonwithPCA}
\begin{figure} 
	\includegraphics[width=1\columnwidth, height=0.45\textheight]{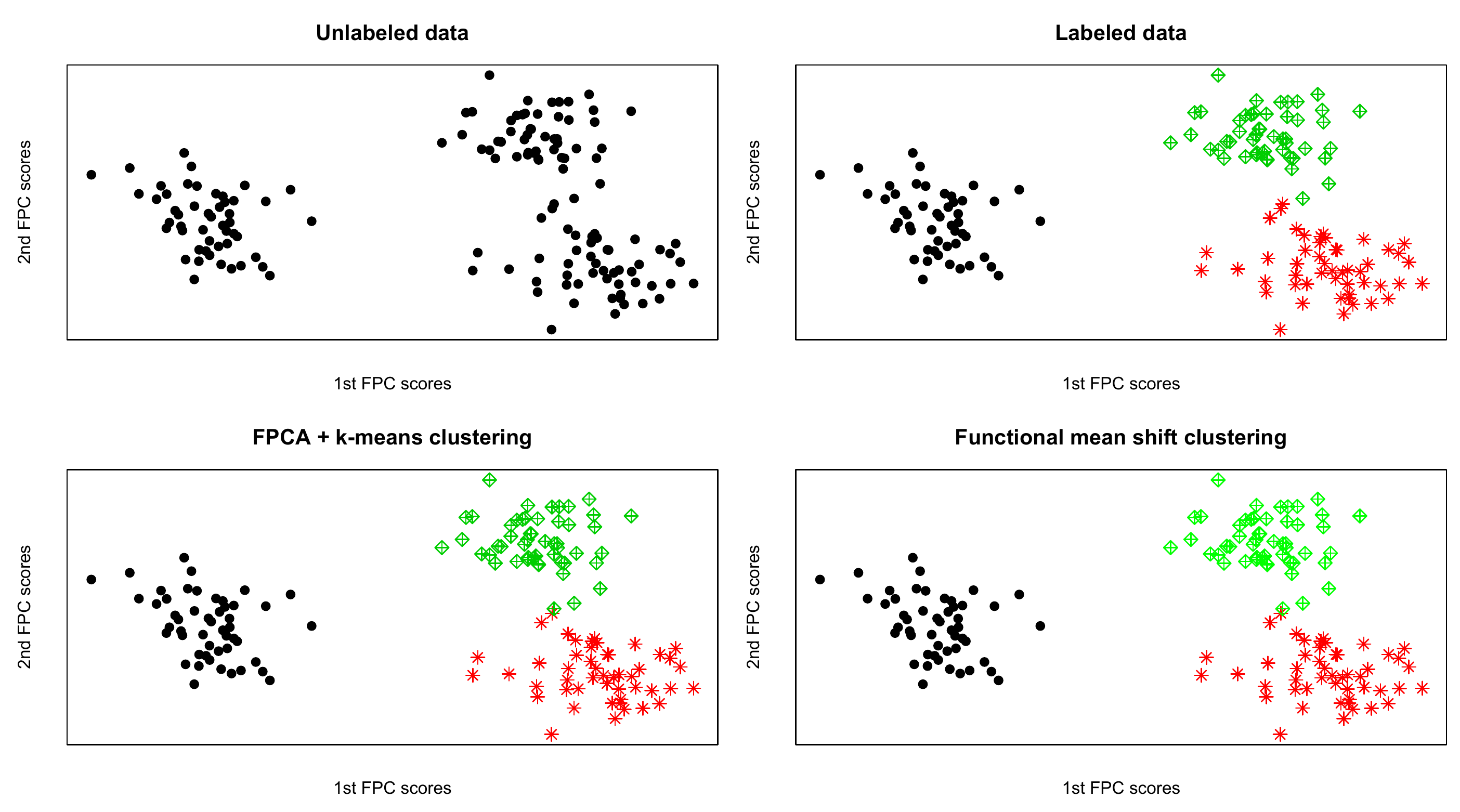}
	\includegraphics[width=1\columnwidth, height=0.2\textheight]{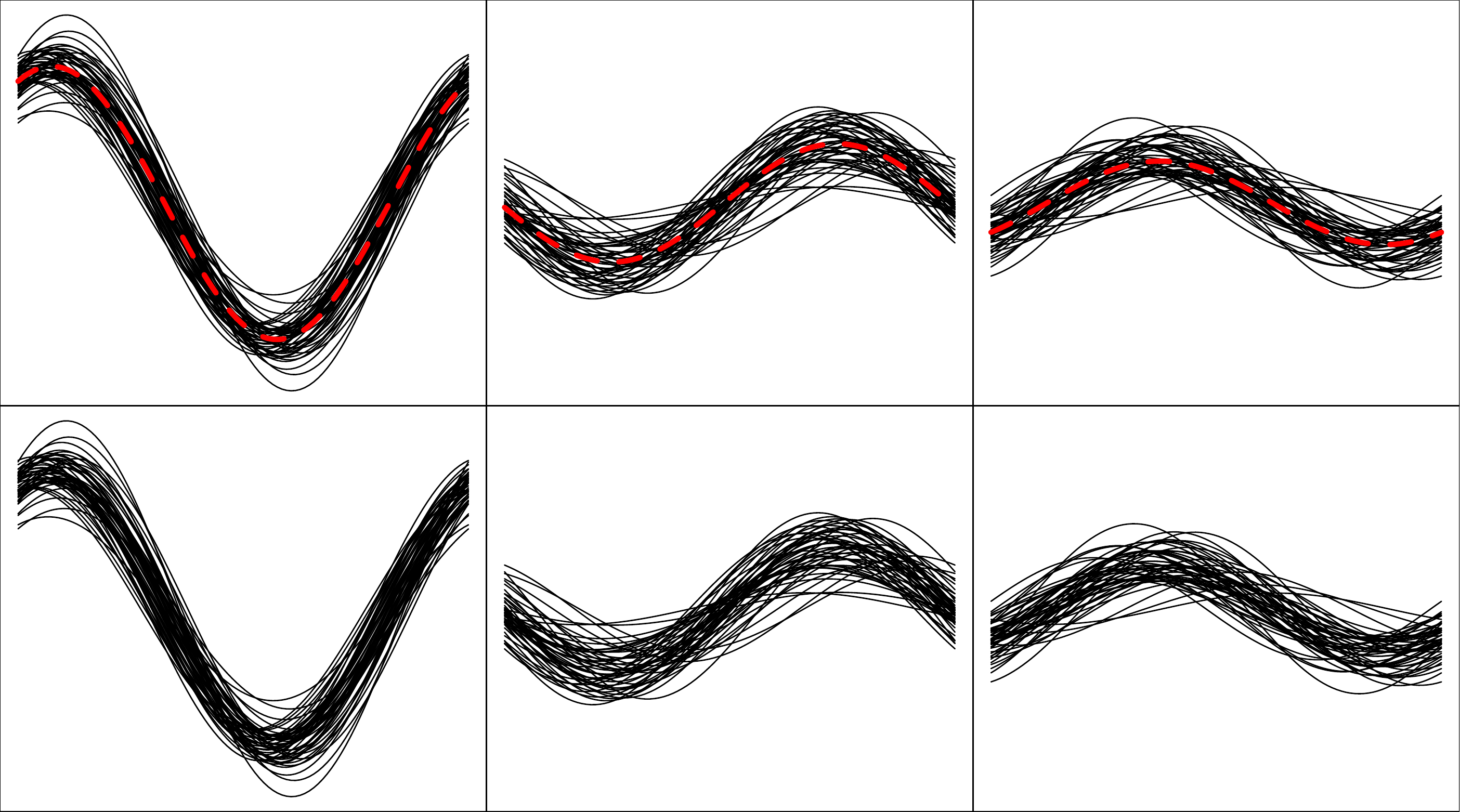}
	\caption{\underline{Top 4 panels:} Projection coefficients on the first two principal components of a set of 150 curves in the span of $\sin(2\pi t)$ and $\cos(2\pi t)$. \underline{Bottom 4 panels:} Clusters recovered by the functional mean-shift algorithm  (top) with modal curves (red dashed lines) and clusters recovered by fPCA/$k$-means (bottom). When the clusters have elliptical shapes, the combined fPCA/$k$-means analysis and the functional mean-shift algorithm both recover the underlying clustering structure of the data.}
	\label{fig:MSvsPCAelliptical}
\end{figure}

For the clustering of a sample of functional data that have a small intrinsic dimensionality (say, $D$), the functional mean-shift algorithm under the $L_2$ distance is expected to perform at least as good as applying $k$-means clustering to the projection coefficients on the first $D$ principal components. If the intrinsic dimensionality is in fact $D$, then the squared distance between two functional data points is
\begin{equation*}
d^2(X_i,X_j)=\|X_i-X_j\|^2 \approx \sum_{k=1}^D (\theta_{i,k} - \theta_{j,k})^2,
\end{equation*}
where $\theta_{i,k}$ is the projection coefficient of $X_i$ on the $k$-th principal component. In this case, clustering using the functional mean-shift algorithm is essentially equivalent to density-based clustering using a kernel density estimate based on the $D$-dimensional projection coefficients. $k$-means tends to produce accurate clusterings when the number of clusters is known and the empirical clusters have elliptical shapes. Hence, roughly speaking, if $D$ is small, the number of clusters is known, and the clusters of principal component scores have elliptical shapes, then the functional mean-shift algorithm with the $L_2$ distance is expected to yield very similar results when compared to $k$-means clustering on the $D$-dimensional projection coefficients of fPCA (Figure \ref{fig:MSvsPCAelliptical}). However, when the intrinsic dimensionality of the functional data is large, the functional mean-shift algorithm has greater flexibility compared to the combined fPCA/$k$-means clustering, in that it can pick up differences between the clusters that may not be evident in the projections onto the lower dimensional space spanned by the first few principal components. Also, as opposed to the fPCA/$k$-means approach, the functional mean-shift algorithm does not require the user to select the number of clusters a priori.

\begin{figure} 
	\includegraphics[width=1\columnwidth, height=0.4\textheight]{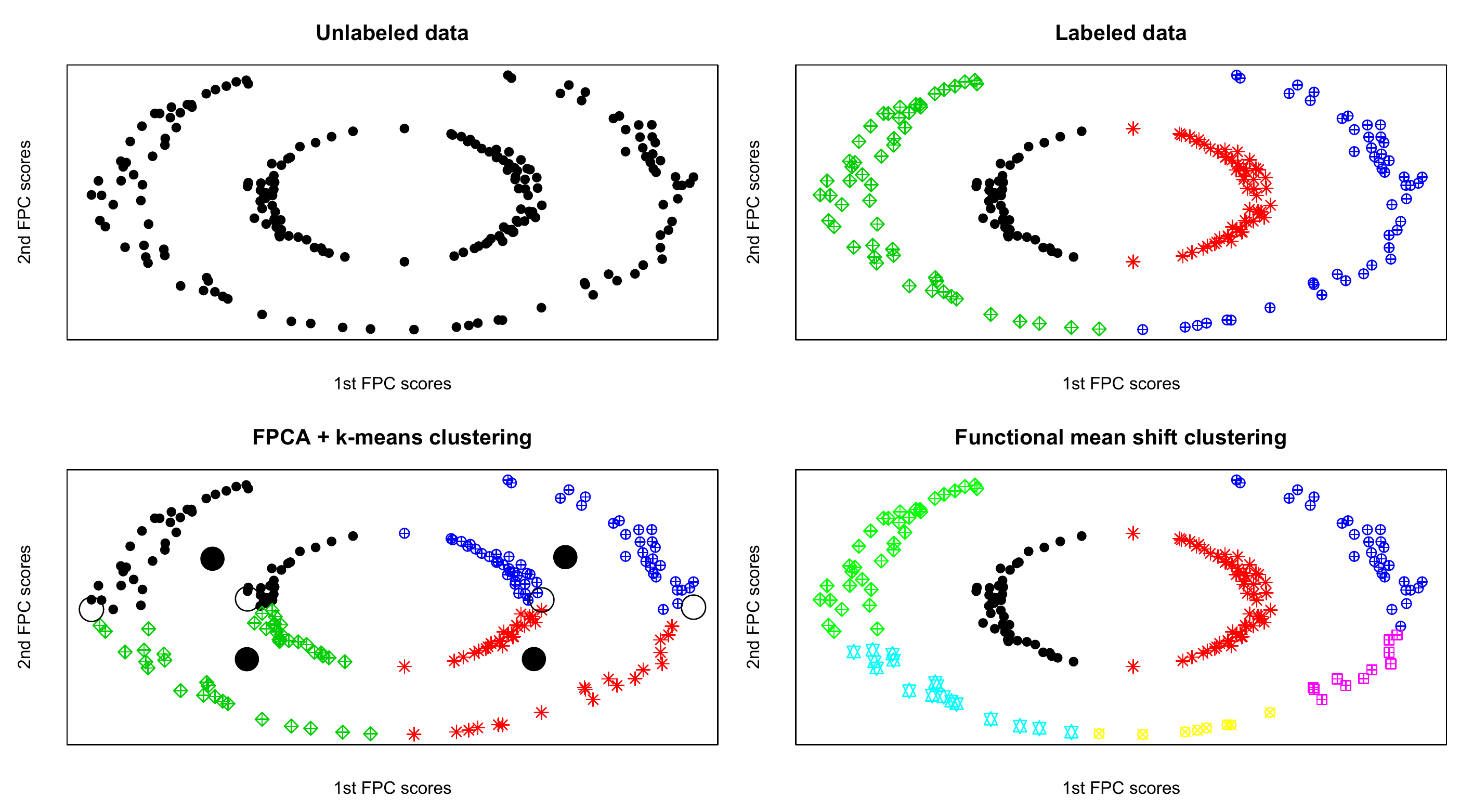}
	\caption{Projection coefficients on the first two principal components of a set of 200 curves in the span of $\sin(2\pi t)$ and $\cos(2\pi t)$. The circular pattern prevents $k$-means clustering from achieving a satisfactory result even when the number of clusters is known a priori and the initial seeds (hollow circles in the bottom left panel) are suitably placed. The large full circles in the bottom left panel represent the final cluster centers obtained with $k$-means. }
	\label{fig:MSvsPCAcircular}
\end{figure}

\begin{figure} 
	\includegraphics[width=1\columnwidth, height=0.2\textheight]{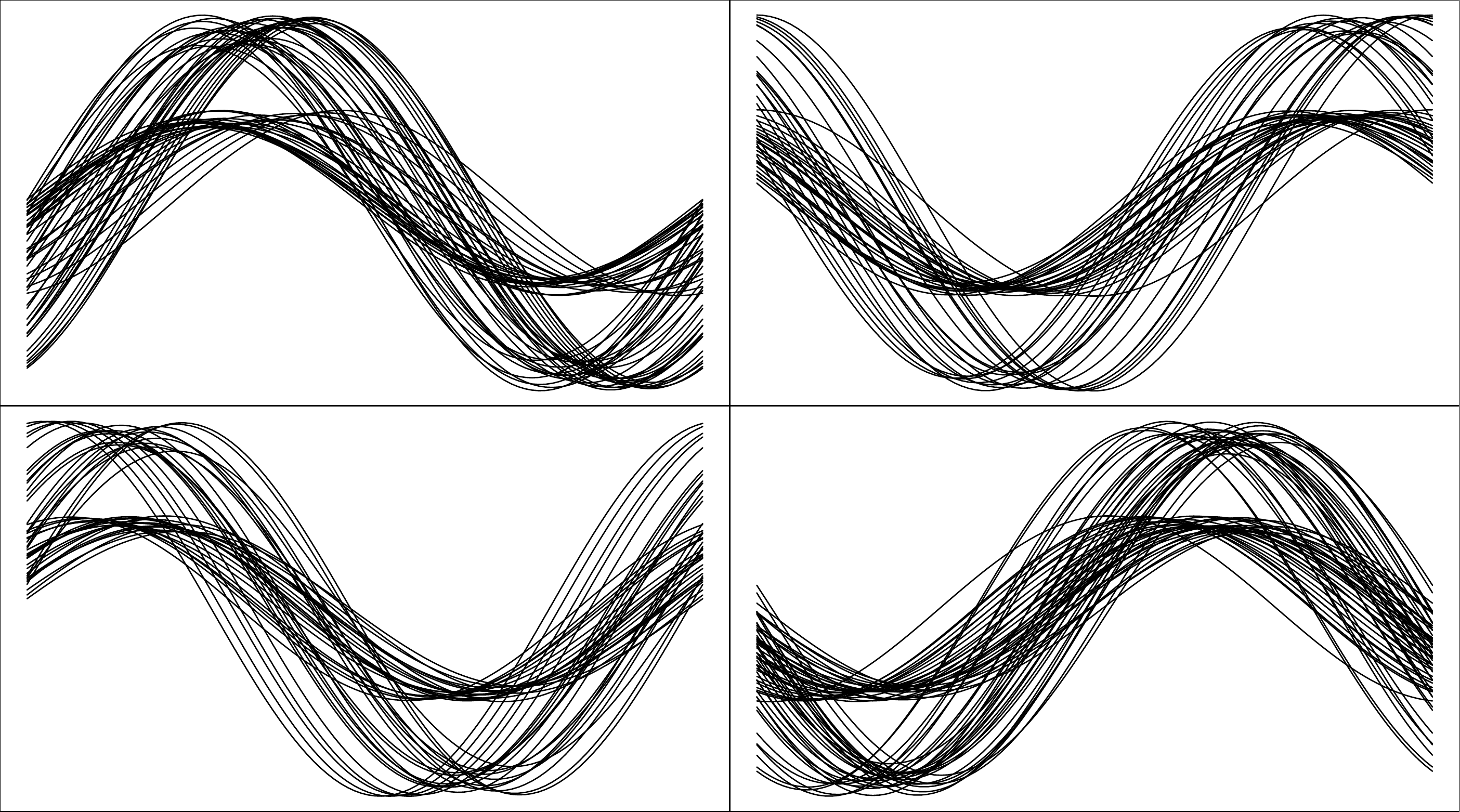}
	\caption{Clusters obtained with the combined PCA/$k$-means analysis.}
	\label{fig:PCAclusters}
\end{figure}

\begin{figure} 
	\includegraphics[width=1\columnwidth, height=0.3\textheight]{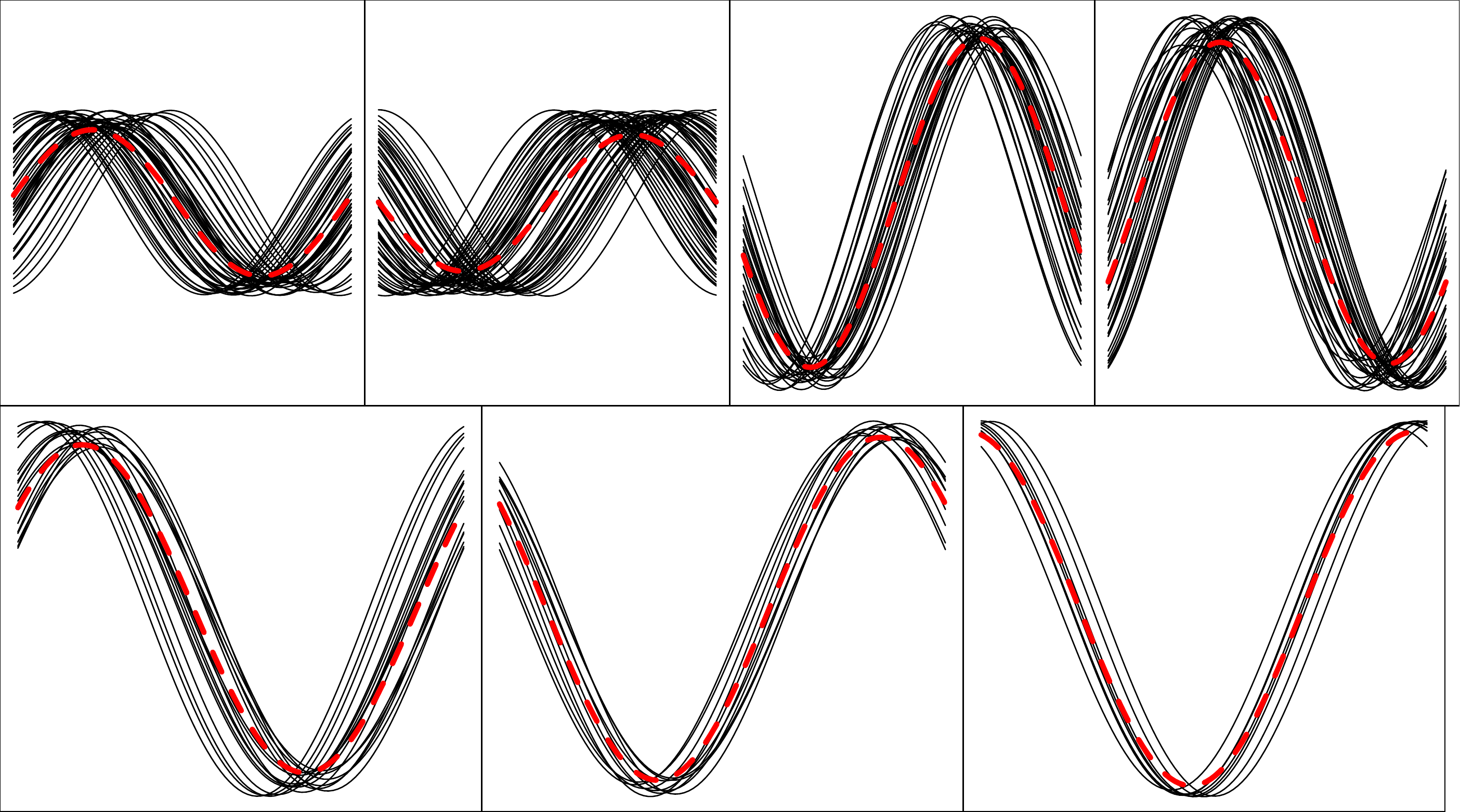}
	\caption{Clusters obtained with the functional mean-shift algorithm. The modal curves are represented by the red dashed lines.}
	\label{fig:MSclusters}
\end{figure}

Interestingly, even in finite dimensional settings there exist situations in which the functional mean-shift algorithm outperforms the combined fPCA/$k$-means analysis. For example, Figure \ref{fig:MSvsPCAcircular} depicts the projections of a set of 200 curves that lie in the span of the functions $\sin(2\pi t)$ and $\cos(2\pi t)$ onto their first two principal components (which explain 100\% of the sample variance). Here, $k$-means is doomed to fail with the circular pattern of Figure \ref{fig:MSvsPCAcircular} even when the number of clusters is known and the seeds are placed suitably. Figure \ref{fig:PCAclusters} depicts the clusters obtained by means of the combined fPCA/$k$-means approach. It is clear that these clusters do not represent the underlying structure of the data.  On the other hand, the functional mean-shift algorithm generates meaningful clusters (Figure \ref{fig:MSclusters}) and effectively summarizes each of the them by its modal curve.
%

\section{Inference on the functional modes}
\label{sec:inference}
In the previous section, we showed that the functional mean-shift algorithm locates the candidate local modes of the estimated surrogate density $\tilde p$, namely a set of points $\tilde{\mathcal{M}}=\{ \tilde \mu_1, \dots, \tilde \mu_r \}$ such that the first order condition 
\begin{equation*}
	\nabla \tilde p_{\mu_j}=0 \iff \tilde p_{\tilde \mu_j}(y)=0 \quad \forall y \in \mathcal{X}
\end{equation*}
is satisfied for $j=1,\dots, r$. We may want to test whether $\tilde \mu_j \in \tilde{\mathcal{M}}$ is a statistically \textit{significant} functional mode of the unknown surrogate density $p$, i.e. we may want to test whether $\tilde \mu_j \in \tilde{\mathcal{M}}$ is a critical point of $p$ and the second Gate\^aux derivative of $p$ is a negative definite operator at $\tilde \mu_j$. Let us assume that the unknown surrogate density is twice Gate\^aux differentiable with second Gate\^aux differential at $x \in \mathcal{X}$ denoted by
\begin{align*}
	p^{(2)}_x :\mathcal{X} \times \mathcal{X} &\to  \mathbb{R} \\
									(y,z) &\mapsto p^{(2)}_x(y,z),
\end{align*}
which is a symmetric continuous bilinear operator (the analog of the Hessian matrix in infinite dimensions). A critical point $\mu_j \in \mathcal{M}$ of the unknown surrogate density $p$ is a local maximum if and only if $p^{(2)}_{\mu_j}$ is negative definite, i.e.
\begin{equation*}
	\sup_{\|y\|=1} p^{(2)}_{\mu_j}(y,y) = \lambda_{\mu_j} < 0.
\end{equation*}
Thus, for $\tilde \mu_j \in \tilde{\mathcal{M}}$, we want to test
\begin{align*}
		\begin{split}
			H_0 : \lambda_{\tilde \mu_j} \geq 0 \quad &\text{or} \quad \nabla p_{\tilde \mu_j} \neq 0 \\
				& \text{vs.} \\
			H_1 : \lambda_{\tilde \mu_j}  < 0 \quad &\text{and} \quad \nabla p_{\tilde \mu_j} = 0,
		\end{split}
\end{align*}
that is
\begin{equation*}
		\begin{split}
			H_0: \tilde \mu_j \text{ is } &\text{not a mode of } p \\
			& \text{ vs. } \\
			H_1: \tilde \mu_j \text{ is } &\text{a mode of } p.
		\end{split}
\end{equation*}
The pointwise nature of the alternative hypothesis on the functional gradient may constitute a problem in the specification of the above hypothesis test. For instance, in the Euclidean setting, the condition $\nabla p_{\tilde \mu_j}=0$ would correspond to a zero (Lebesgue) measure set. At least in the Euclidean setting, one possibility is to recast the hypothesis test as follows (\citealp{genovese2013nonparametric}):
\begin{align}
	\label{eq:hypothesistest}
		\begin{split}
			H_0 &: \neg H_1 \\
				& \text{vs.} \\
			H_1 : \lambda_{\tilde \mu_j}  < 0 \quad &\text{and} \quad \|\nabla p_{\tilde \mu_j}\| < \delta\,
		\end{split}
\end{align}
for some $\delta >0$, so that we are instead testing the hypothesis that $\tilde \mu_j$ is an \textit{approximate} mode of $p$. The sample splitting procedure introduced in the test of \cite{genovese2013nonparametric} is such that constraint on the gradient has essentially no effect in \eqref{eq:hypothesistest} (we will later return to this point in Remark \ref{remark:remarkontest}). In the following, we consider a similar test that is adapted to the functional case. We focus on
\begin{align}
	\label{eq:hypothesistesteffective}
		\begin{split}
			H_0 &: \lambda_{\tilde \mu_j} \geq 0\\
				& \text{vs.} \\
			H_1 &: \lambda_{\tilde \mu_j}  < 0.
		\end{split}
\end{align}
A natural test statistic for \eqref{eq:hypothesistesteffective} is $\tilde \lambda_{\hat \mu_j}=\sup_{\|y\|=1} \tilde p^{(2)}_{\hat \mu_j}(y,y)$. Lemma \ref{lemma:secondGateauxdifferential} gives the explicit form of the second Gate\^aux differential of the estimated surrogate density at an arbitrary point $x \in \mathcal{X}$. The explicit form of the test statistic $\tilde \lambda_x$ is then derived in Lemma \ref{lemma:test statistics}.
\begin{lemma}[Second Gate\^aux differential of the estimated surrogate density]
\label{lemma:secondGateauxdifferential}
The second Gate\^aux differential of the estimated surrogate density $\tilde p$ at $x \in \mathcal{X}$ evaluated at $(y,z) \in \mathcal{X} \times \mathcal{X}$ is
\begin{align*}
	\tilde p_x^{(2)}(y,z) &= - C w_G(\mathcal{S}) \sum_{X \in \mathcal{S}} \frac{1}{h^2(X)} \Bigg[\frac{1}{h(X)} K_h'(X,x) \frac{\langle X-x,y \rangle \langle X-x,z \rangle}{\|X-x\|} \\
					&+ K_h(X,x) \langle y, z \rangle \Bigg].	
\end{align*}
\end{lemma}
\begin{lemma}[Test statistic for the second order optimality condition]
	\label{lemma:test statistics}
	We have
	\begin{align*}
		\tilde \lambda_x &= \sup_{\|y\|=1} \tilde p^{(2)}(y,y) = Cw_G(\mathcal{S}) \left[ 2 \left\| \sum_{X \in \mathcal{S}} \frac{1}{h^3(X)}K'_h(X,x)\frac{X-x}{\|X-x\|}\right\| -  \right.\\\
			&  \left. \sum_{X \in \mathcal{S}} \frac{1}{h^2(X)}\left[ \frac{1}{h(X)}K'_h(X,x)\left( \|X-x\| + \|X-x\|^{-1} \right) + K_h(X,x)\right]\right].
	\end{align*}
\end{lemma}
\begin{remark}
Notice that when the profiles $g$ and $k$ satisfy the conditions of Remark \ref{remark:conditionsonprofiles}, $\lim_{t \to 0^+} \frac{k'(t)}{t}$
exists and it is finite. Thus, we can safely set $\frac{k'(0)}{0}=\lim_{t \to 0^+} \frac{k'(t)}{t}$ in the expressions involved in the two previous Lemmata whenever needed.
\end{remark}
We are now in the position to describe our testing procedure. Suppose for simplicity that the sample $\mathcal{S}$ contains an even number of elements, say $2n$, and that we use a non-adaptive bandwidth parameter $h(X)=h>0$ for all $X \in \mathcal{S}$.\\

\textbf{Stage 1:} First, we divide the sample in two subsamples of size $n$. We apply the functional mean-shift algorithm on the first subsample, denoted $X_1, \dots, X_n$, in order to determine the set $\tilde{\mathcal{M}}=\{\tilde \mu_1, \dots, \tilde \mu_r \}$ of candidate local modes of $\tilde p$, which is an estimate of the set $\mathcal{M}=\{\mu_1,\dots,\mu_R\}$ of local modes of the unknown surrogate density $p$. This step usually requires using the first subsample $X_1, \dots, X_n$ to determine the value of the bandwidth $h$ for the functional mean-shift algorithm.\\

\textbf{Stage 2:} Next, we use the second subsample, denoted $Y_1, \dots, Y_n$, to compute the test statistic $\tilde \lambda_{\tilde \mu_j}$ for $j=1,\dots, r$. For $b=1,\dots,B$ we sample with replacement $n$ elements of the second subsample, $Y_{1,b}^*,\dots,Y_{n,b}^*$ from the empirical distribution of $Y_1,\dots,Y_n$, $P_n(y)=\frac{1}{n} \sum_{i=1}^n \delta_{Y_i}(y)$. For $b=1,\dots,B$ we compute the test statistic $\tilde \lambda^*_{\tilde \mu_j,b}$ using the bootstrap sample $Y^*_{1,b},\dots,Y^*_{n,b}$ and we thus construct a $1-\alpha$ bootstrap confidence interval for $\lambda_{\tilde \mu_j}$, $\mathcal{C}_{n,1-\alpha}(\lambda_{\tilde \mu_j})$ where $\alpha \in (0,1)$. Finally, if $\sup \left \{ \lambda \in \mathcal{C}_{n,1-\alpha}(\lambda_{\tilde \mu_j})\right\}<0$, we reject the null hypothesis that $\tilde \mu_j$ is not a local mode of the unknown surrogate density $p$.

\begin{figure} 
	\includegraphics[width=1\columnwidth, height=0.3\textheight]{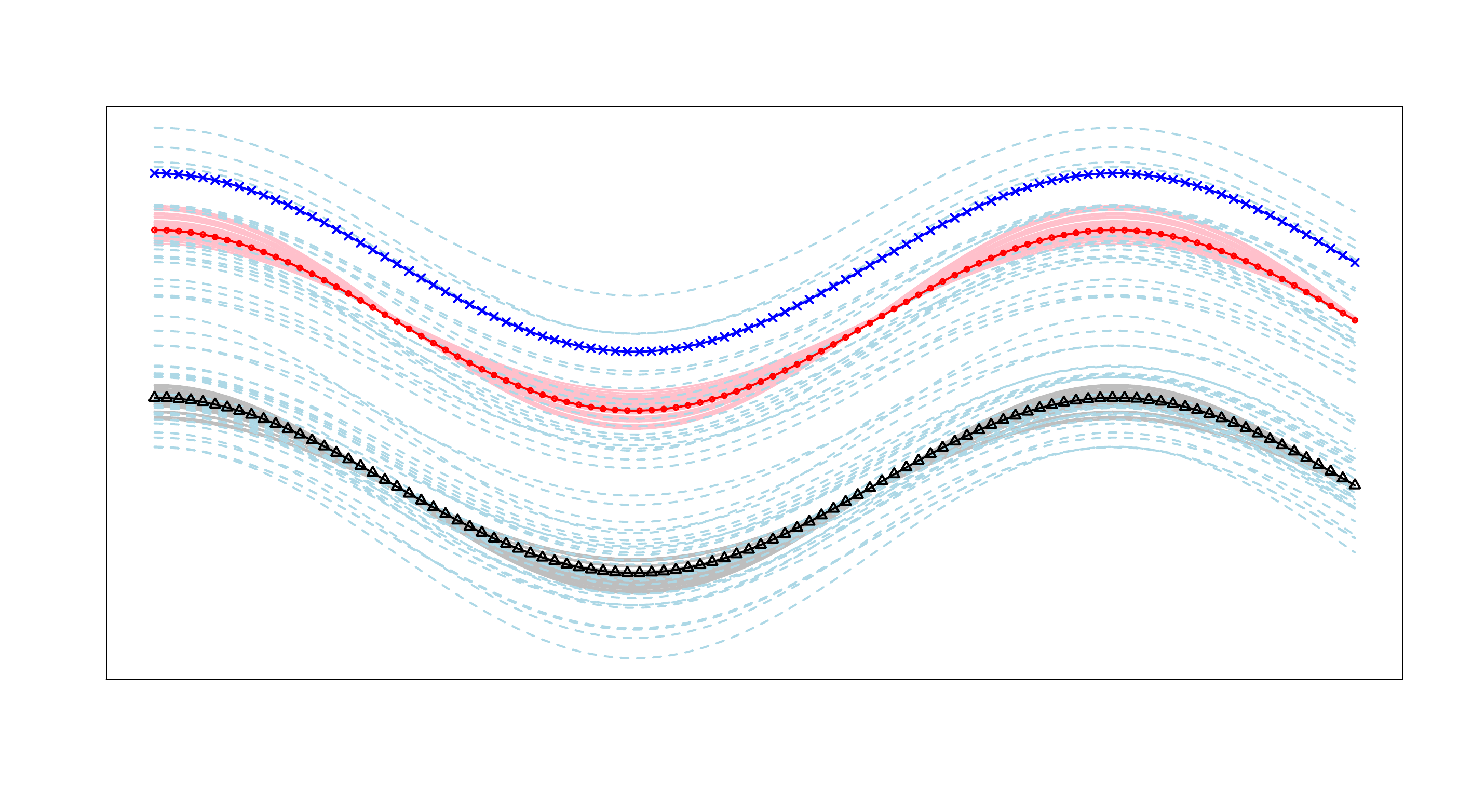}
	\includegraphics[width=1\columnwidth, height=0.3\textheight]{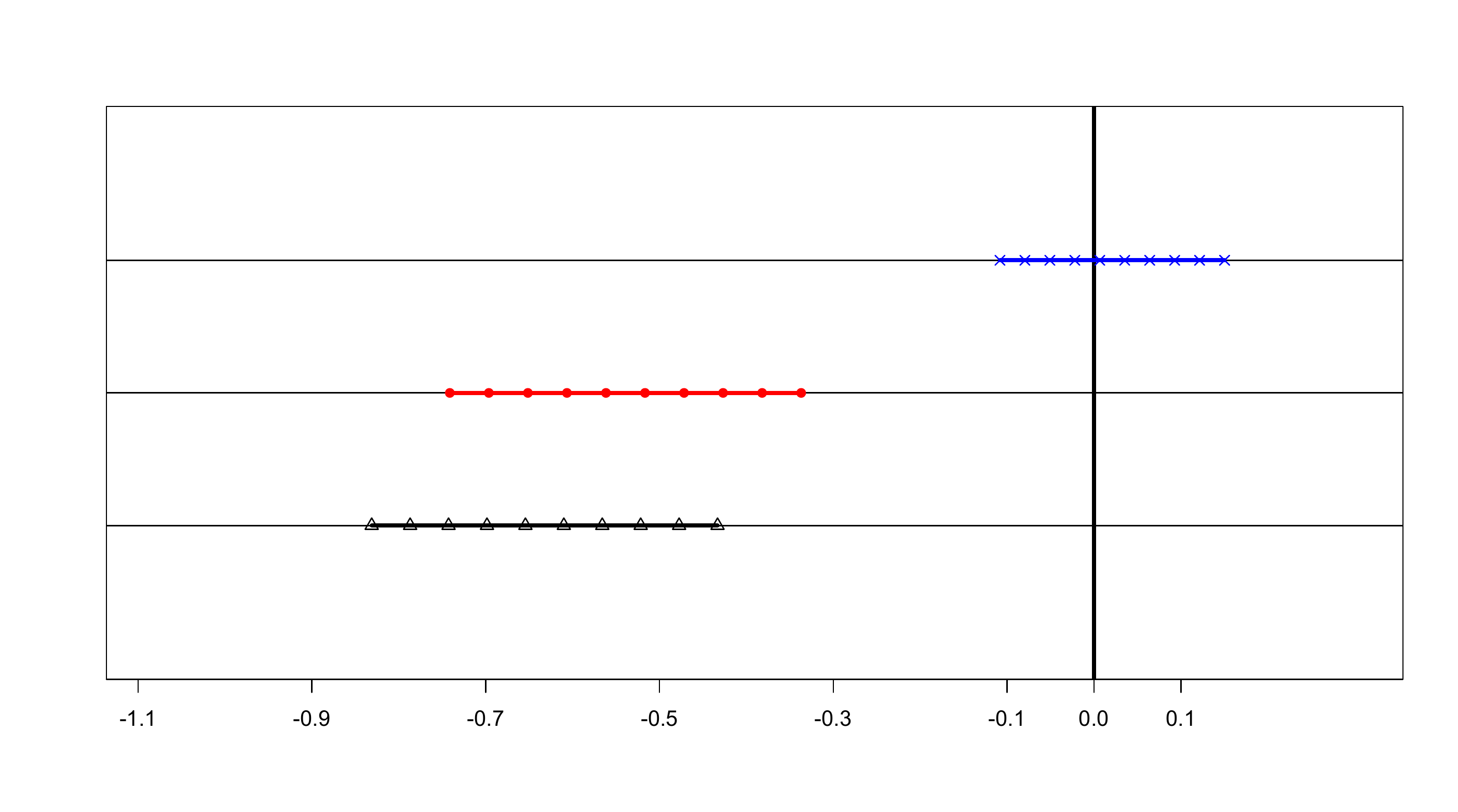}
	\caption{\underline{Top panel:} Simulated data. The red and the black curves (top and bottom bundles of continuous lines) are representative of two similar, but distinct, signals. The blue curves (dashed lines) are clutter curves. The highlighted curves are the 3 candidate functional local modes identified by the functional mean-shift algorithm, $\tilde \mu_X$, $\tilde \mu_C$, and $\tilde \mu_Y$. \underline{Bottom panel:} Simultaneous approximate 95\% confidence intervals for $\lambda_{\tilde \mu_X}$, $\lambda_{\tilde \mu_C}$, and $\lambda_{\tilde \mu_Y}$ using $B=1000$ bootstrap replications. Only the confidence intervals associated to the candidate modal curves $\tilde \mu_X$ and $\tilde \mu_Y$ entirely lie to the left of 0. Therefore, as expected, only $\tilde \mu_X$ and $\tilde \mu_Y$ appear to correspond to significant functional local modes.}
	\label{fig:simulateddata}
\end{figure}

By setting $\alpha'=\alpha / \ell$, we can use the same procedure to construct a confidence rectangle for $\ell \in \{1, \dots, r \}$ of the $\lambda_{\tilde \mu_j}$'s simultaneously. This rectangle takes the form $\mathcal{C}_{n,1-\alpha} \left(\lambda_{\tilde \mu_{j_1}}, \dots,\lambda_{\tilde \mu_{j_\ell}} \right)=\mathcal{C}_{n,1-\alpha'}\left(\lambda_{\tilde \mu_{j_1}}\right) \times \dots \times \mathcal{C}_{n,1-\alpha'}\left(\lambda_{\tilde \mu_{j_\ell}}\right)$. Note that the purpose of sample splitting in Stage 1 is to assure the validity of the confidence intervals. The simultaneous test algorithm is summarized in Figure \ref{fig:test}.
\begin{remark}
\label{remark:remarkontest}
Because each candidate local mode $\tilde \mu_j \in \tilde{\mathcal{M}}$ is such that $\nabla \tilde p_{\tilde \mu_j}=0$ in the first stage, the estimated gradient is likely to be null also in the second stage. Thus, the constraint $\| \nabla p_{\tilde \mu_j} \| < \delta$ has a negligible effect in \eqref{eq:hypothesistest}.
\end{remark}
%
%
\begin{figure}
\fbox{\parbox{\textwidth}{
\begin{center}
	\textbf{Local functional mode set testing algorithm}\\
\end{center}
\noindent \textbf{Input:} sample $\mathcal{S}$ with $\#(\mathcal{S})=2n$, coverage level $1-\alpha$\\
\textbf{Output:} a set $\tilde{\mathcal{M}}^S \subset \mathcal{X}$ of significant local modes at the approximate level $1-\alpha$

\begin{enumerate}
	\item split the sample in two halves, $X_1, \dots, X_n$ and $Y_1, \dots, Y_n$
	\item apply the mean-shift algorithm on $X_1, \dots, X_n$ to find the set of candidate local modes $\tilde{\mathcal{M}}=\{ \tilde \mu_1, \dots, \tilde \mu_r\}$
	\item for each $j=1, \dots, r$ and for $b=1, \dots, B$ do
		\begin{itemize}
			\item resample with replacement $n$ elements $Y^*_{1,b}, \dots, Y^*_{n,b}$
			\item compute $\tilde \lambda^*_{\tilde \mu_j,b}$
		\end{itemize}
	\item for each $j=1,\dots,r$ construct the $1-\alpha/r$ level confidence interval $\mathcal{C}_{n,1-\alpha/r}(\lambda_{\tilde \mu_j})$ using the empirical quantiles of $\tilde \lambda^*_{\tilde \mu_j,1}, \dots, \tilde \lambda^*_{\tilde \mu_j,B}$.
	\item for each $j=1, \dots, r$ check the condition $\sup \left\{ \lambda \in \mathcal{C}_{n,1-\alpha/r}(\lambda_{\tilde \mu_j}) \right\} < 0$
	\item set $\tilde{\mathcal{M}^S}=\left \{ \tilde \mu_j \in \tilde{\mathcal{M}}: \sup \left\{ \lambda \in \mathcal{C}_{n,1-\alpha/r}(\lambda_{\tilde \mu_j}) \right\} < 0 \right\}$
	\item output $\tilde{\mathcal{M}^S}$.
\end{enumerate}
}}
\caption{The local functional mode set testing algorithm. Sample spitting assures the validity of the test.}
\label{fig:test}
\end{figure}
Let us now illustrate the above testing procedure by means of a simple simulation. We randomly draw 150 curves from these three groups:
\begin{itemize}
	\item $X(t) = \eta_{X} \cos \left( 5\pi /2 \cdot t \right)$
	\item $Y(t) = 3 + \eta_{Y} \cos \left( 5\pi /2 \cdot t \right)$
	\item $C(t) = \gamma + 3\cdot \mathbbm{1}_{\{U>0.5\}}  + \cos \left( 5\pi /2 \cdot t \right)$
\end{itemize}
where $\eta_{X} \overset{\text{d}}=\eta_{Y} \sim \mathcal{N}(\mu=1, \sigma=0.1)$, $\gamma \sim \mathcal{N}(\mu=0, \sigma=0.8)$, $U \sim \text{Unif}(0,1)$, $t \in [0,1]$, and all the random variables involved are independent of each other. The probability of drawing a curve from a given group is the same for each group and equal to $1/3$. In this simulation, the $X$'s and the $Y$'s are thought of as realizations of two distinct \textit{signals}, whereas the $C$'s are \textit{clutter} curves that correspond to a version of these signals that is subject to a vertical perturbation. On the basis of the data generating process, it makes sense to expect the existence of a modal curve $\mu_X$ for the $X$'s and a distinct modal curve $\mu_Y$ for the $Y$'s. However, the clutter curves are likely to generate some spurious local mode (at least for small enough bandwidths) when one estimates $\mu_X$ and $\mu_Y$ using the functional mean-shift algorithm. The simulated curves are presented in the top panel of Figure \ref{fig:simulateddata} (the $X$'s is grey, the $Y$'s in red and the $C$'s in light blue). The functional mean-shift identifies 3 candidate functional modal curves $\tilde \mu_X$, $\tilde \mu_C$, and $\tilde \mu_Y$, with the asymmetric truncated Gaussian kernel (top panel of Figure \ref{fig:simulateddata}). Inside the kernel, we set $d$ to be the $L_2$ distance and the bandwidth is set to $h=7.5$ (approximately the 41th percentile of the $L_2$ distances among the curves in the first subsample). As expected, only the local modes corresponding to the $X$'s and the $Y$'s are significant on the basis of the local mode testing algorithm of Figure \ref{fig:test} (in this simulation, we set the number of bootstrap replications $B$ equal to 1000). The output of the testing procedure and the 95\% simultaneous approximate confidence intervals for $\lambda_{\tilde \mu_X}$, $\lambda_{\tilde \mu_C}$, and $\lambda_{\tilde \mu_Y}$ are summarized in the bottom panel of Figure \ref{fig:simulateddata}.
%

\section{Application: spike sorting}
\label{sec:neural}
In this section, we describe how the functional mean-shift algorithm can be used to cluster a set of curves that correspond to neural activity.

\begin{figure} 
	\includegraphics[width=1\columnwidth, height=0.3\textheight]{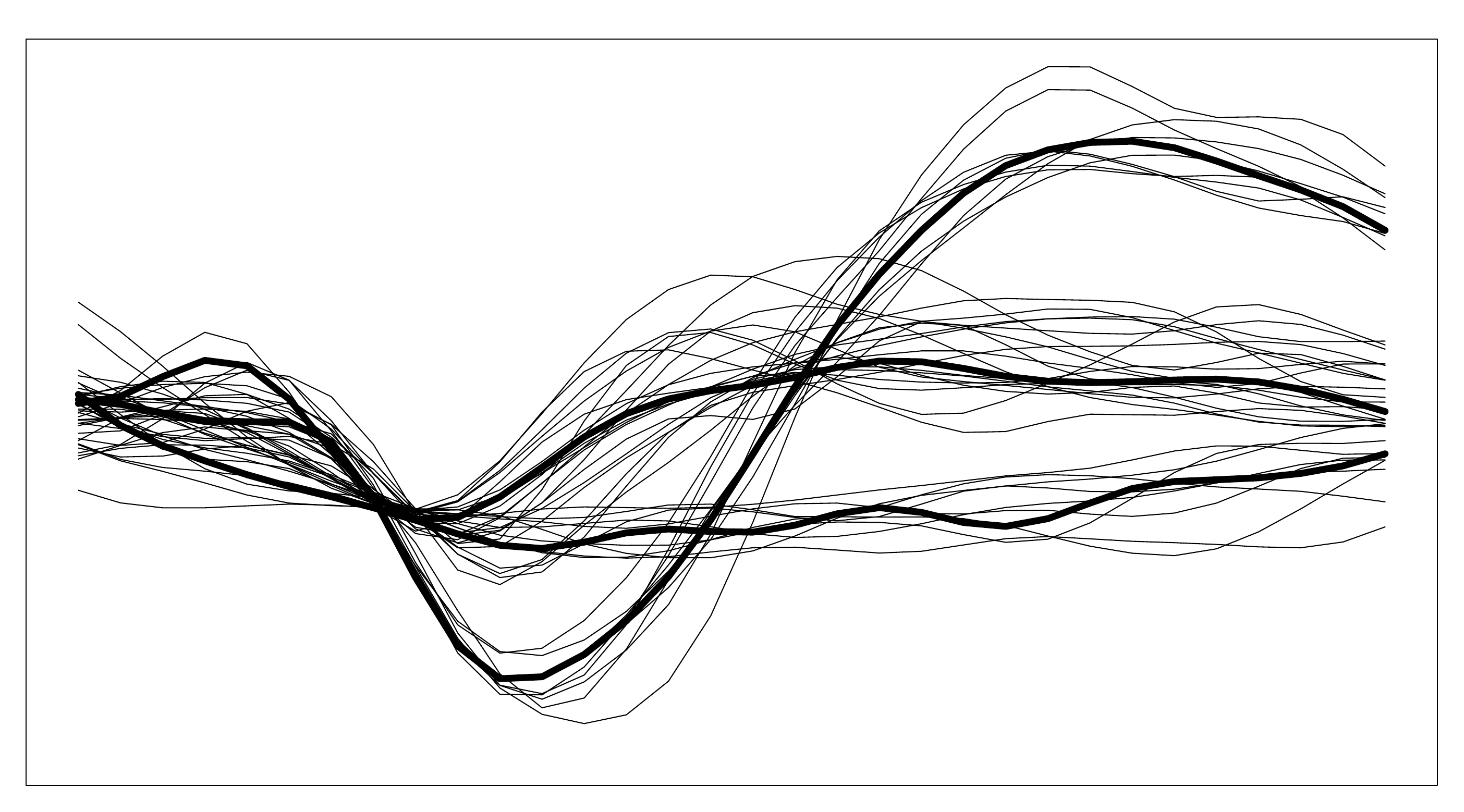}
	\caption{Plot of the 40 neural activity curves. Three curves are highlighted to show typical trajectories. On the basis of this figure, one expects that the curves can be clustered in 3 homogeneous groups, each characterized by its own modal trajectory.}
	\label{fig:neuralcurvesdataset}
\end{figure}

The available data, displayed in Figure \ref{fig:neuralcurvesdataset}, represent a subset of 40 recordings of neurons over time which come from a behavioral experiment performed at the Andrew Schwartz motorlab (\url{http://motorlab.neurobio.pitt.edu/index.php}) on a macaque monkey (the authors thank Andrew Schwartz, Val\'erie Ventura and Sonia Todorova for providing the data). The monkey performs a center-out and out-center target reaching task with 26 targets in a virtual 3D environment. The curves of Figure \ref{fig:neuralcurvesdataset} show the voltage of neurons versus the times recorded at electrodes (32 equidistant time points per curve; time is standardized between 0 and 1). The recorded neural activity consists of all the action potentials detected above a channel-specific threshold on a 96-channel Utah array implanted in the primary motor cortex.

The goal is to perform {\em spike sorting}, i.e. clustering the curves in a set of distinct homogeneous groups. Each cluster of curves is then thought to correspond to the activity of a single neuron since each neuron tends to have its own characteristic curve (or {\em spike}). An analysis of (a larger set of) these curves can also be found in \cite{conformalexplore2013lei}, who cluster the curves following the \textit{conformal prediction} approach (\citealp{vovk2009line}).

\begin{figure} 
	\includegraphics[width=1\columnwidth, height=0.3\textheight]{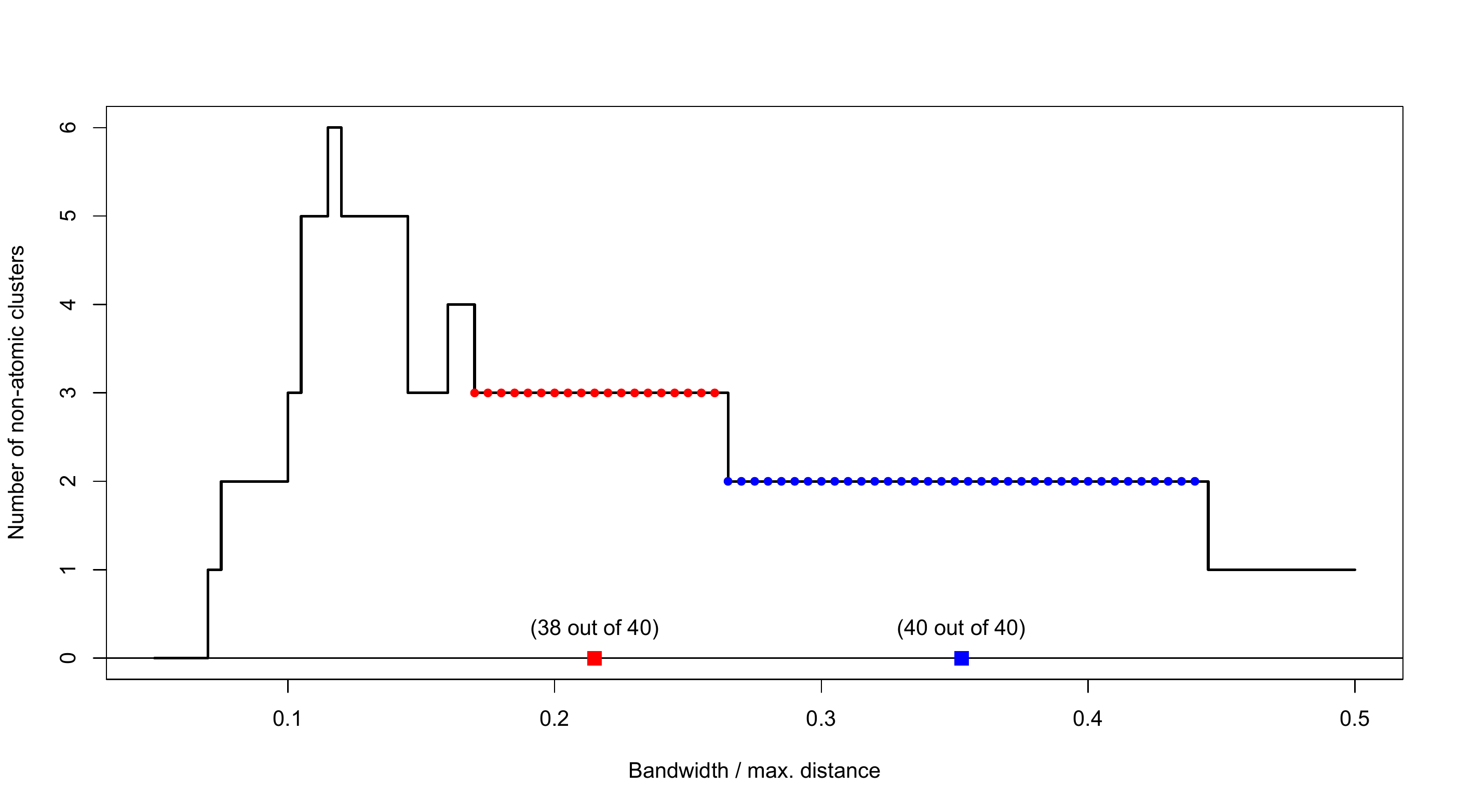}
	\caption{Number of non-atomic clusters of neural activity curves as a function of the bandwidth parameter, $h$. The number of non-atomic clusters stabilizes in the 2 highlighted ranges. The candidate bandwidth parameters are then set to be the midpoint of the 2 relevant ranges. The numbers in parentheses above the candidate bandwidths denote the number of sample curves that are clustered in non-atomic clusters using the corresponding candidate bandwidth.}
	\label{fig:nuberofmodes_neural}
\end{figure}

\begin{figure} 
	\includegraphics[width=1\columnwidth, height=0.3\textheight]{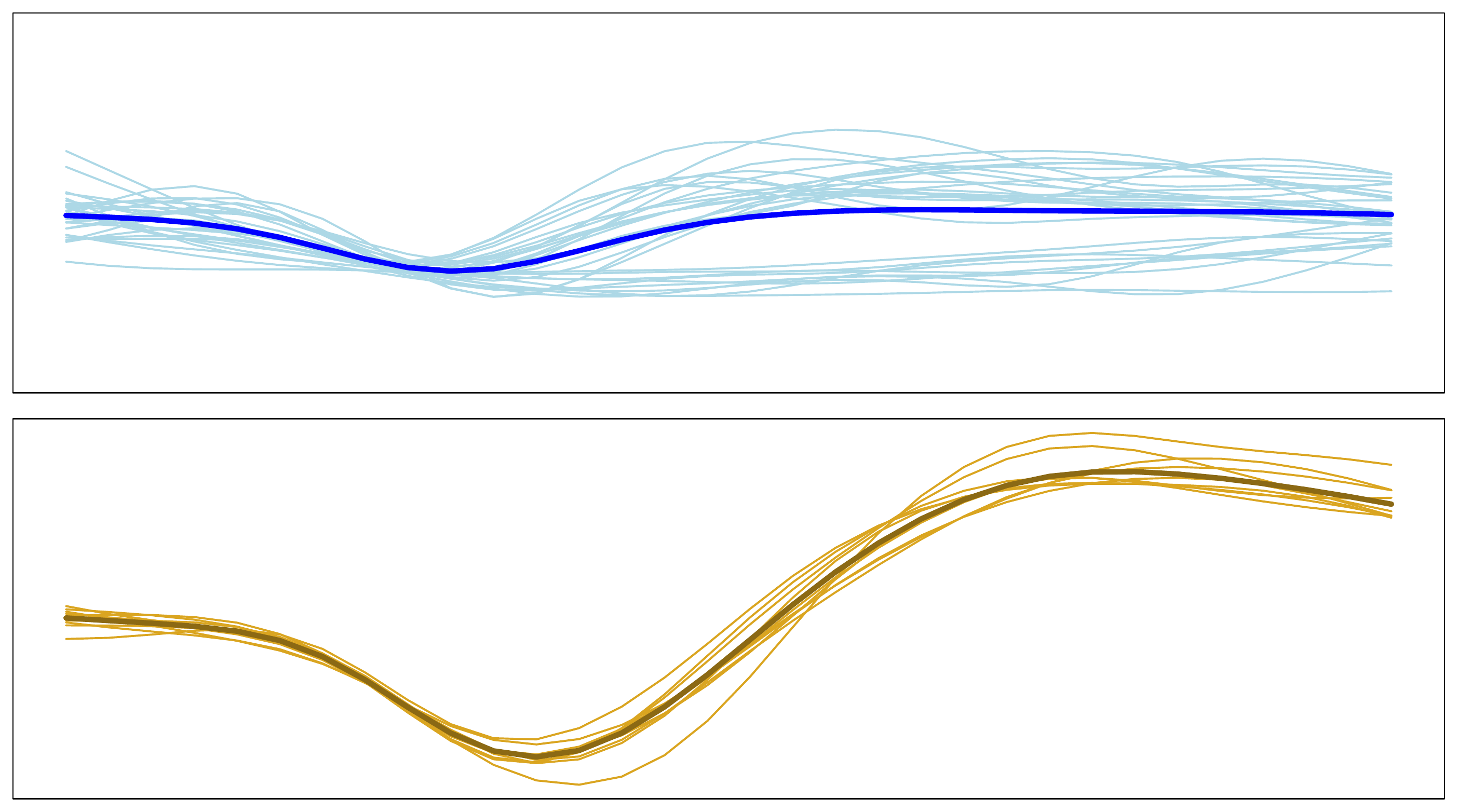}
	\caption{Output of the functional mean-shift algorithm on the 40 neural activity curves with the largest candidate bandwidth, $h_2$. The plots are on the same scale. Each panel represents an individual cluster of neural activity curves and the bold line is the estimated modal curve for that cluster. With this choice of the bandwidth, the functional mean-shift algorithm detects that there are two bundles of curves that share a similar shape (top blue cluster) and a third cluster corresponding to curves with a more pronounced sigmoid shape (bottom yellow cluster).}
	\label{fig:MSresultsNeural2}
\end{figure}

\begin{figure} 
	\includegraphics[width=1\columnwidth, height=0.3\textheight]{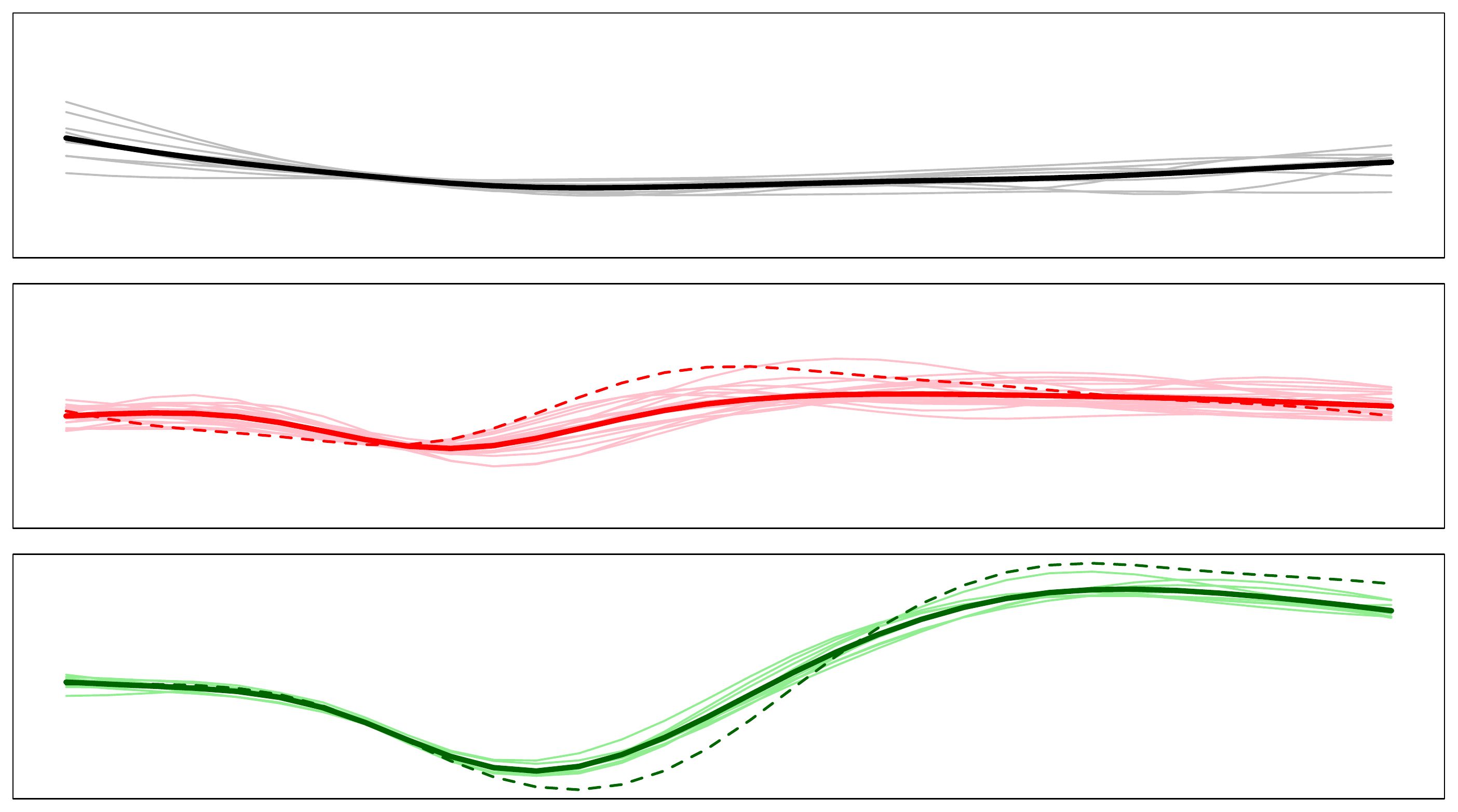}
	\caption{Output of the functional mean-shift algorithm on the 40 neural activity curves with the smallest candidate bandiwidth, $h_1$. The plots are on the same scale. Each panel represents an individual cluster of neural activity curves and the bold line is the estimated modal curve for that cluster. From the plots, it appears that the choice of the bandwidth is just right: the functional mean-shift algorithm detects three internally homogeneous clusters. With this choice of the bandwidth, the algorithm identifies 2 atomic clusters (potential outliers) which correspond to the dashed curves in the two lower panels.}
	\label{fig:MSresultsNeural3}
\end{figure}

On the basis of Figure \ref{fig:neuralcurvesdataset}, one may expect to find 3 clusters, each summarized by a corresponding modal curve. We set $d$ to be the distance associated with the Sobolev space $H_{1}$ of square integrable functions on the standard unit interval with square integrable first weak derivative, i.e.
\begin{equation*}
	d(x,y)=\|x-y\|_{L_2} + \|x' - y' \|_{L_2}.
\end{equation*}
This distance is associated to the inner product
\begin{equation*}
	\langle x, y \rangle_{H_1} = \langle x, y \rangle_{L_2} + \langle x', y' \rangle_{L_2}.
\end{equation*}
The curves are smoothed using local quadratic polynomials and their first derivatives are obtained directly from the local polynomial fit. We apply the functional mean-shift on this dataset of curves using the asymmetric truncated Gaussian kernel. A fixed bandwidth parameter is chosen using a heuristic based on the plot of the number of non-atomic clusters (that is, clusters of sample curves containing more than a single curve) as a function of bandwidth (Figure \ref{fig:nuberofmodes_neural}). In particular, we repeatedly run the functional mean-shift using an increasing sequence of bandwidths (100 equally separated values between $5\%$ and $50\%$ of the largest observed distance between the sample curves). The candidate bandwidths are chosen as the midpoints of the ranges where the number of clusters stabilizes. For these data, this heuristic identifies $h_1=0.2150\cdot \max_{X,X' \in \mathcal{S}}d(X,X')$ and $h_2=0.3525 \cdot \max_{X,X' \in \mathcal{S}}d(X,X')$ as the two candidate bandwidths.
The output of the mean-shift algorithm is displayed in Figures \ref{fig:MSresultsNeural2} and \ref{fig:MSresultsNeural3}. In this `multi-bandwidth' analysis, the largest candidate bandwidth, $h_2$, produces a coarse clustering of the sample curves: from Figure \ref{fig:MSresultsNeural2}, we see that the two most similar bundles of curves are clustered together in a unique cluster. At this resolution, the algorithm recognizes that these two bundles have a more similar shape, which is different from the sigmoid shape of the third bundle. The smallest candidate bandwidth, $h_1$, produces a clustering of the sample curves that is just right (Figure \ref{fig:MSresultsNeural3}). At this resolution, the functional mean-shift algorithm also identifies two atomic clusters (dashed lines in the lower two panels of Figure \ref{fig:MSresultsNeural3}) which, in light of Remark \ref{remark:outliers}, could be considered potential outliers.
%
%

\section{Application: signature forgery}
\label{sec:signatures}
\begin{figure} 
	\includegraphics[width=1\columnwidth, height=0.3\textheight]{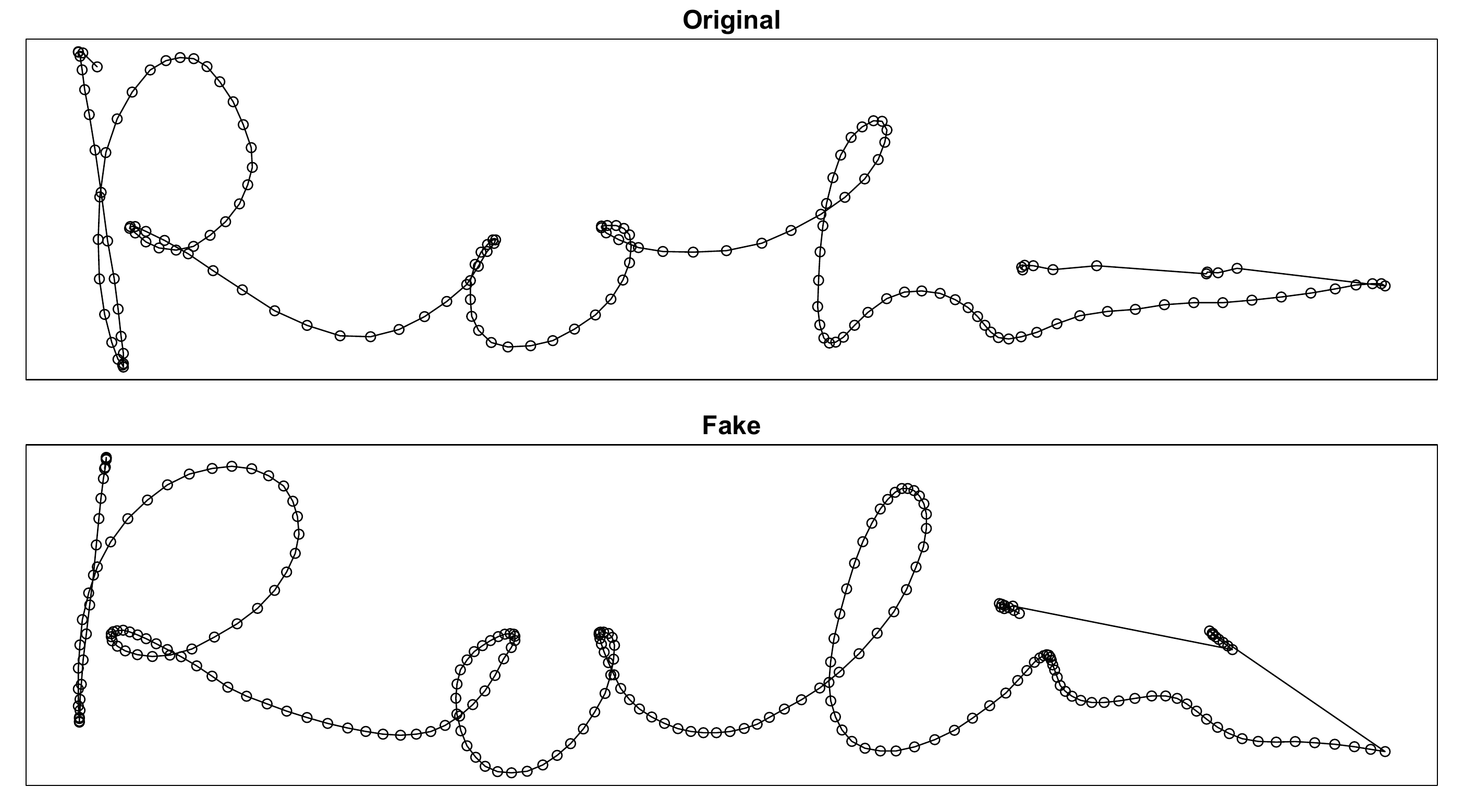}
	\caption{Examples of signatures from a sample of 40 SVC 2004 signatures (20 original, 20 fake). The upper panel displays an example of the owner's signature, while the bottom panel contains an example of a forged signature.}
	\label{fig:signatures}
\end{figure}

Suppose that a dataset of $n$ similar signatures, $\mathcal{S}=\{X_1, \dots, X_n\}$, is available to us. Each observed signature $X \in \mathcal{S}$ can be thought of as a (discretized version of a) curve in $\mathbb{R}^2$ of the type
\begin{equation*}
	t \mapsto X(t)=(x(t),y(t)),
\end{equation*}
where, for simplicity, we can take $t \in [0,1]$. To simplify, we may identify each signature in the dataset with its {\em tangential acceleration}, i.e. the real valued function
\begin{equation}
	\label{eq:tangentialacceleration}
	t \mapsto S(t) = \left(x''(t),y''(t) \right) \cdot \frac{\left(x'(t),y'(t)\right)}{\| \left(x'(t),y'(t) \right)\|_2},
\end{equation}
which captures the amount of acceleration along the line tangent to the signature $X$ at time $t$.

From now on we identify the sample of signatures with the i.i.d. sample of tangential accelerations $\mathcal{S}=(S_1, S_2, \dots, S_n)$, $S_1 \sim P$. It is reasonable to imagine that if all the signatures in $\mathcal{S}$ are produced by the same author (the owner of the signature), then the common distribution $P$ of the $S$'s is `unimodal'. In fact, in this case we might expect the distribution of the tangential accelerations to concentrate around a central tangential acceleration curve in the space of the $S$'s (the `typical' tangential acceleration of the owner's signature); the scattering of the other tangential accelerations around this central curve would correspond to the random variation among different instances of an original signature. However, if some of the signatures in $\mathcal{S}$ are instead forged by one or more skilled fakers whose forgeries appear to be good replicates of the original owner's signature to the naked eye, but whose tangential acceleration curves are sufficiently different from the tangential acceleration curves of the owner, then we may expect $P$ to be `multimodal'.

Suppose that we do not know exactly how many different authors produced the signatures in $\mathcal{S}$, i.e. we suspect that some signatures could be forged versions of the owner's signature. Running the functional mean-shift algorithm on $\mathcal{S}$ would output of a set $\tilde{\mathcal{M}}=\{\tilde \mu_1, \dots, \tilde \mu_r\}$ of modal tangential accelerations (corresponding to local modes of $\tilde p$, the estimated surrogate density of $P$) and would assign each observed tangential acceleration $S \in \mathcal{S}$ to the closest mode in $\tilde{\mathcal{M}}$, thus allowing us to divide the sample signatures in $r$ distinct clusters. If the functional mean-shift algorithm detects a unique modal signature, and therefore all the observed signatures are grouped in a unique cluster, we may conclude that there is no evidence (at least on the basis of the tangential accelerations) that there exist forged signatures in $\mathcal{S}$. However, if $\tilde{\mathcal{M}}$ contains more than a single modal curve ($r>1$), and thus the functional mean-shift algorithm partitions $\mathcal{S}$ in more than a single cluster, then we have empirical evidence that the sample $\mathcal{S}$ may contain signatures both from the owner and (potentially) other authors. In this case, we may conclude that some signatures in $\mathcal{S}$ are forged versions of the owner's original signature.

In this section, we test the functional mean-shift on a set of 40 signatures from a sample catalog of the SVC 2004 Signature Verification Competition (\url{http://www.cse.ust.hk/svc2004/download.html}, Sample Data, User 1). While it is known that in this sample there are 20 original owner's signatures (corresponding to the first 20 curves in the dataset) and 20 fake signatures (corresponding to the curves numbered from 21 to 40 in the dataset), we pretend that we do not know this information a priori. If the two groups of signatures are sufficiently separated in the space of the tangential accelerations, then the functional mean-shift should detect at least two modal curves. Our analysis represents an unsupervised counterpart of the analysis of \cite{geenenssignatures}, who develops a nonparametric functional classifier for these data.

\begin{figure} 
	\includegraphics[width=1\columnwidth, height=0.3\textheight]{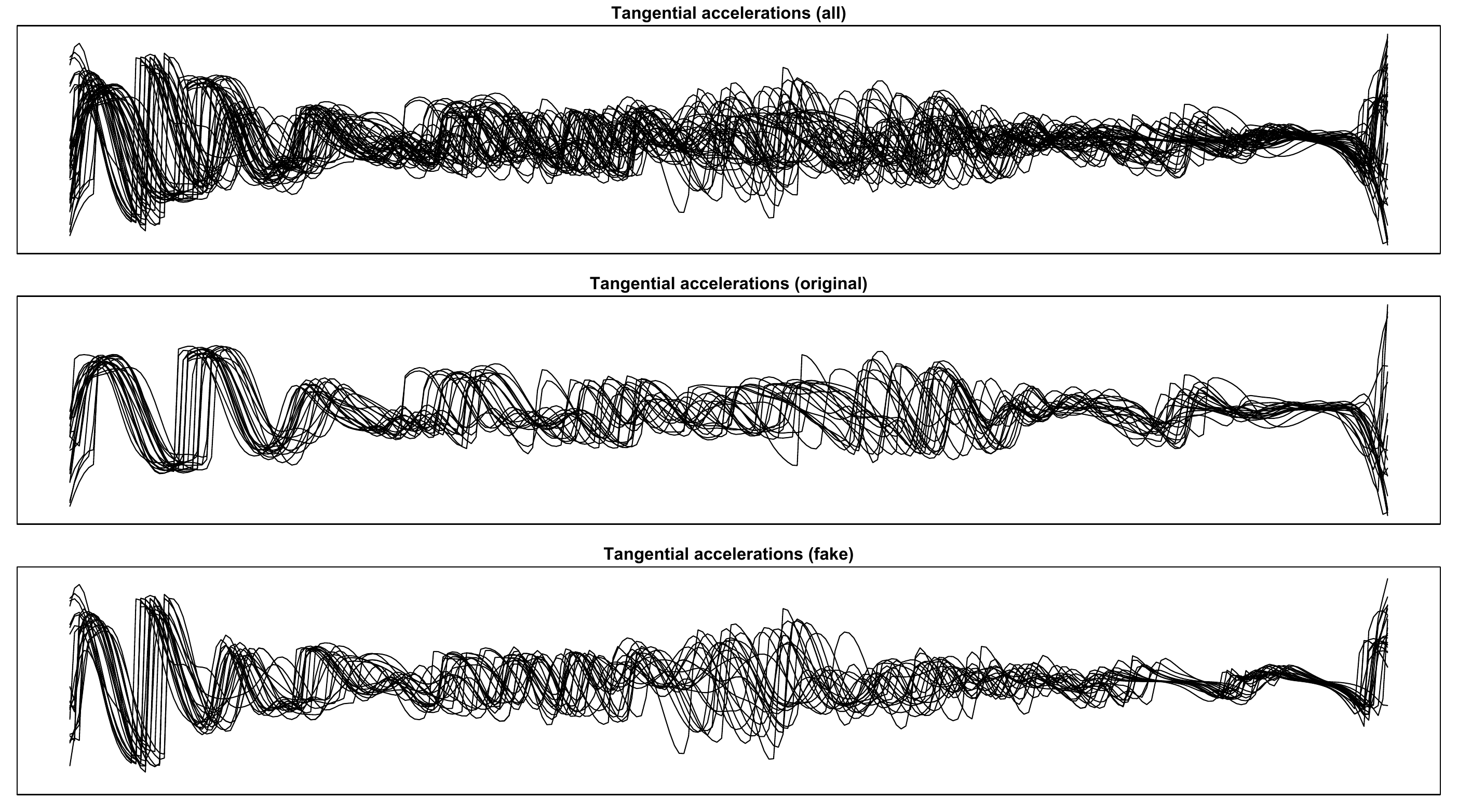}
	\caption{Smooth tangential accelerations obtained from the balanced dataset of 40 SVC 2004 signatures. The top panel displays the smooth tangential accelerations of all the 40 signatures together, whereas the middle and the bottom panels display the smooth tangential accelerations for the original and for the fake signatures, respectively.
}
	\label{fig:tangentialaccelerations}
\end{figure}

Figure \ref{fig:signatures} displays two signatures (one original, one fake) from the SVC 2004 sample. It is not immediately evident to the naked eye that the two signatures are not produced by the same author. Our first step is to smooth the $x$ and $y$ components of the 40 signatures in the sample with a locally quadratic polynomial to get a smooth representation of $x'$, $x''$, $y'$ and $y''$. In particular, we use the \verb+locpoly+ function of the \verb+KernSmooth+ R package to perform the local polynomial smoothing and the function \verb+dpill+ of the same package to select the bandwidth parameter. With these smooth estimates of the derivatives, we then obtain the smooth tangential acceleration curve of equation \eqref{eq:tangentialacceleration} for each of the 40 signatures in the dataset which is then normalized so that it has unit $L_2$ norm. Figure \ref{fig:tangentialaccelerations} depicts the normalized smooth tangential accelerations obtained from the smooth representation of $x'$, $x''$, $y'$ and $y''$ and equation \eqref{eq:tangentialacceleration}. 

\begin{figure} 
	\includegraphics[width=1\columnwidth, height=0.3\textheight]{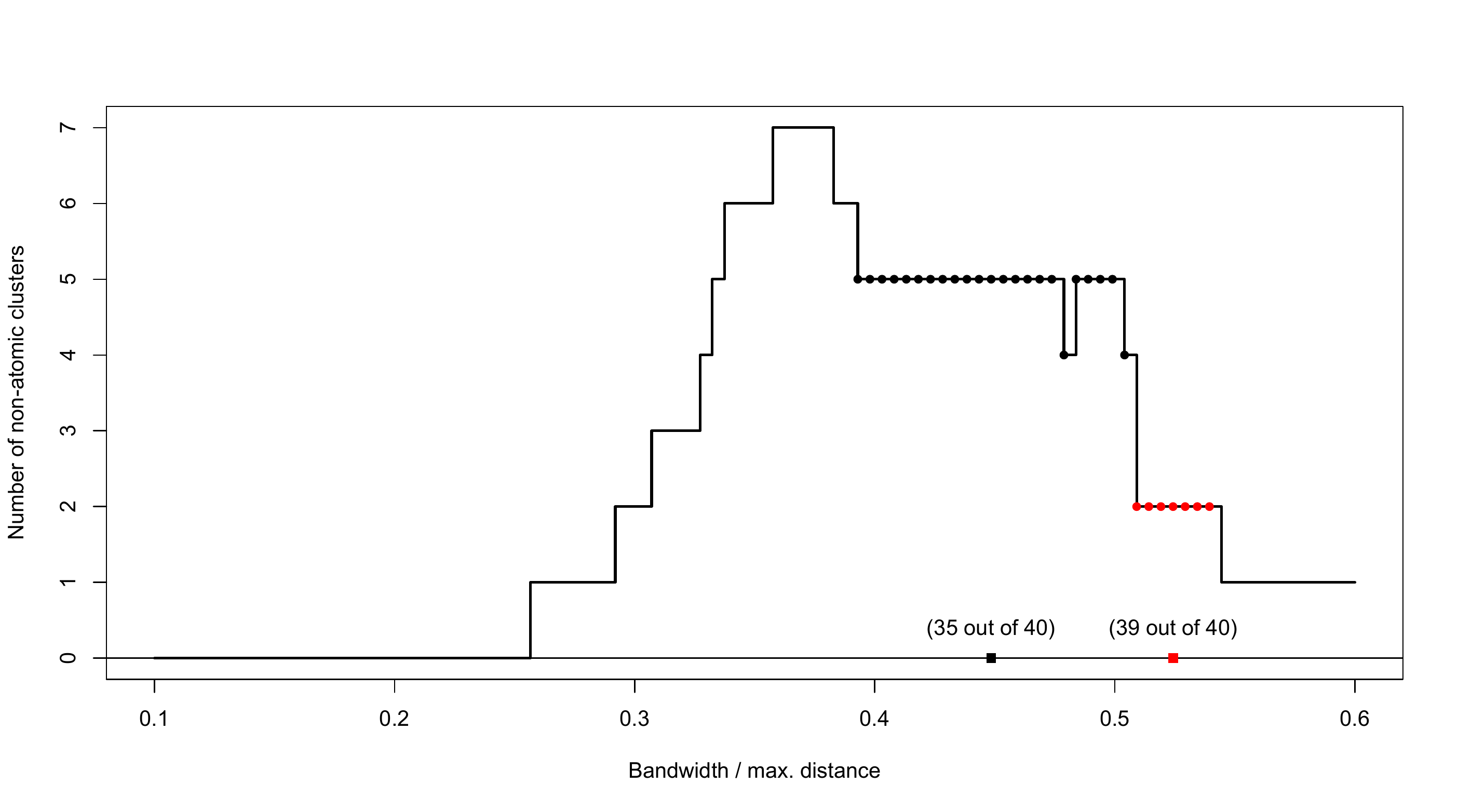}
	\caption{Number of non-atomic clusters as a function of the bandwidth parameter, $h$. The number of non-atomic clusters stabilizes in the 2 highlighted ranges. The candidate bandwidth parameters are then set to be the midpoint of these 2 relevant ranges. The number in parentheses above the candidate bandwidths denote the number of sample curves that are clustered in non-atomic clusters using the corresponding candidate bandwidth.}
	\label{fig:nuberofmodes_signatures}
\end{figure}

We apply the functional mean-shift algorithm with the asymmetric truncated Gaussian profile on the smooth tangential acceleration curves. Within the kernel, $d$ is set to be the distance induced by the $L_2$ norm (as in \citealp{geenenssignatures}), and the same heuristic of Section \ref{sec:neural} is used to select two candidate bandwidths, $h_1=0.4485 \cdot \max_{X,X'}d(X,X')$ and $h_2=0.5242 \cdot \max_{X,X'}d(X,X')$. The plot of non-atomic clusters as a function of bandwidth for the tangential acceleration curves is depicted in Figure \ref{fig:nuberofmodes_signatures}.

\begin{figure} 
	\includegraphics[width=1\columnwidth, height=0.6\textheight]{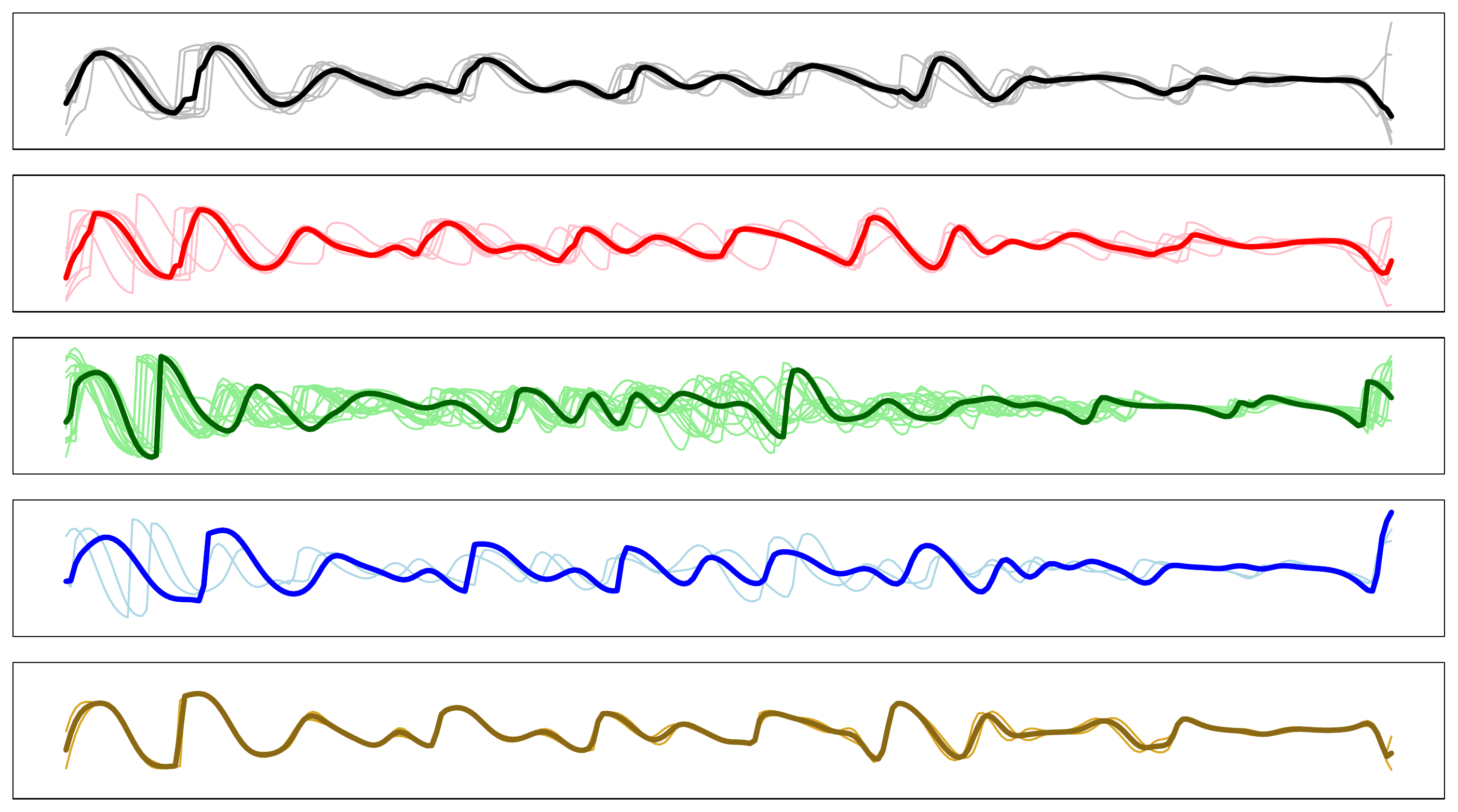}
	\caption{Output of the functional mean-shift algorithm on the 40 smooth tangential acceleration curves when $h=h_1$. The plots are on the same scale. Each panel represents an individual cluster of tangential acceleration curves and the bold line is the estimated modal curve for that cluster. The candidate local mode of the black cluster (top panel) contains 8 original signatures and appears very similar to that of the red (6 original, 1 fake), blue (2 original, 1 fake) and yellow (2 original) clusters (respectively, the second, the fourth and the fifth plots from the top). This similarity suggests that $h_1$ is producing a very fine clustering of the curves. The green cluster (mid plot) contains 15 of the 20 fake signatures. The remaining 5 atomic clusters are not shown in this figure.
}
	\label{fig:MSresults}
\end{figure}

\begin{figure} 
	\includegraphics[width=1\columnwidth, height=0.3\textheight]{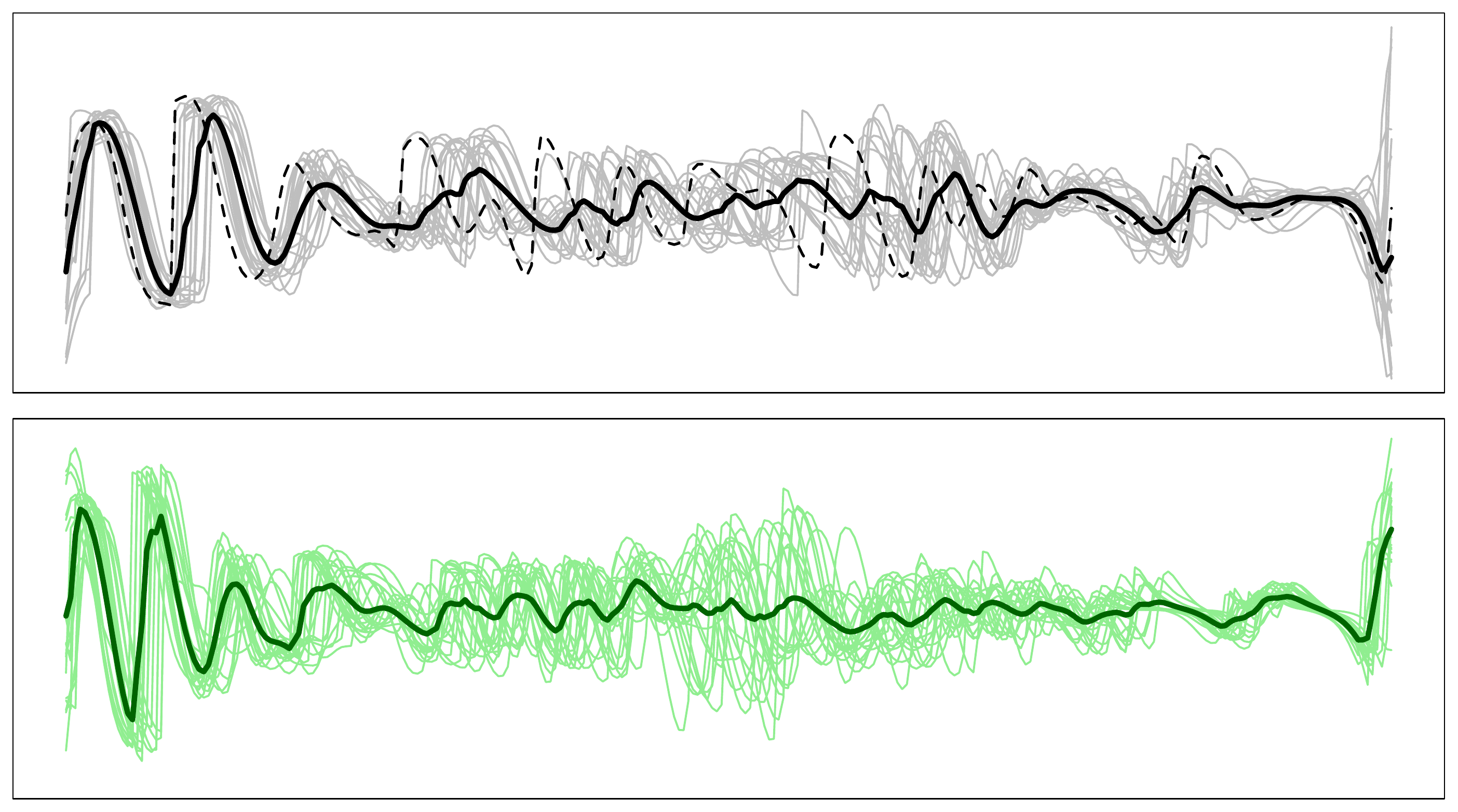}
	\caption{Output of the functional mean-shift algorithm on the 40 smooth tangential acceleration curves when $h=h_2$. The plots are on the same scale. Each panel represents an individual cluster of tangential acceleration curves and the bold line is the estimated modal curve for that cluster. The black cluster (top) contains 19 of the 20 original signatures, whereas the green cluster (bottom) contains the 20 fake signatures. With this choice of the bandwidth, the functional mean-shift algorithm achieves a perfect separation between original and fake signatures. The algorithm finds an atomic cluster (dashed curve in the top panel, corresponding to the first original signature in the dataset), which is therefore flagged as a potential outlier. 
}
	\label{fig:MSresults2}
\end{figure}

Figure \ref{fig:MSresults} displays the output of the functional mean-shift on the smooth tangential acceleration curves when $h=h_1$ and Figure \ref{fig:MSresults2} shows the output of the algorithm when $h=h_2$. The functional mean-shift algorithm finds 5 clusters with $h=h_1$: the top cluster (black) contains 8 original signatures and appears very similar to the second, the fourth, and the fifth clusters (red, blue and yellow). These clusters contain 6 original and 1 fake signature (red), 2  original and 1 fake signature (blue), and 2 original signatures (yellow), respectively. The middle cluster (green) contains 15 of the 20 fake signatures. Overall, it seems that the clustering may be too fine. Figure \ref{fig:MSresults2} displays the output of the functional mean-shift algorithm when $h=h_2$. This time, the algorithm produces two internally homogeneous clusters which appear to meaningfully summarize the structure of the data: the top cluster (black) contains 19 of the 20 original signatures and the second cluster (green) contains all the 20 fake signatures.
%

\section{Discussion}
\label{sec:discussion}
The mean-shift algorithm is an iterative algorithm that corresponds to a form of adaptive gradient ascent. The mean-shift algorithm has been extensively studied for scalar and vector data, but a counterpart of this algorithm for infinite-dimensional data has not been devised. In this paper, we introduce the \textit{functional mean-shift algorithm}, a functional version of the mean-shift algorithm of \cite{fukunagahostetlermeanshift} and \cite{chengmeanshift} which is designed to deal with data that are intrinsically infinite-dimensional.

The functional mean-shift algorithm expands the applicability of the mean-shift algorithm well beyond the familiar Euclidean case. In Section \ref{sec:meanshiftfunctionspaces}, we show that the functional mean-shift algorithm corresponds to a form of adaptive gradient ascent on the estimated surrogate density of random variables that are valued in an infinite-dimensional Hilbert space: for practitioners, this fact establishes a reassuring formal analogy with the standard mean-shift algorithm for scalar and vector data. In Section \ref{sec:inference}, we provide a bootstrap-based simultaneous significance test for the local modes in infinite dimensions. We illustrate by means of a simulation that the test allows us to infer which of the candidate local modes identified by the functional mean-shift algorithm (and therefore which of the clusters) are significant and correspond to real features of the data generating process. In Sections \ref{sec:neural} and \ref{sec:signatures}, we consider two examples of application which demonstrate the effectiveness of the algorithm for the task of clustering smooth curves.

A strength of the algorithm lies in the fact that it does not require the application of a dimension reduction procedure on the sample curves, nor does it require the user to choose the number of clusters a priori (in contrast to PCA-based clustering and $k$-means). Furthermore, the functional mean-shift algorithm is a flexible algorithm because it allows the user to tune the distance function and therefore to choose implicitly in which space to embed the data. 

As it is the case with scalar or vector data, the effectiveness of the functional mean-shift algorithm comes at the cost of being able to properly select the bandwidth parameter $h$. The selection of the bandwidth is a crucial step in any analysis based on the mean-shift algorithm: a poor selection of the bandwidth can drive the algorithm to produce puzzling output. Several automatic and data-driven bandwidth selectors have been proposed for the mean-shift algorithm in the Euclidean setting, some of which reduce to the problem of optimally choosing the bandwidth parameter for the estimation of the underlying probability density (see \citealp{park1990comparison} and \citealp{sheather1991reliable} who develop optimal \textit{plug-in} bandwidth selectors or, among others, \citealp{faraway1990bootstrap} for a bootstrap-based bandwidth selector). This is arguably not the optimal strategy to select the bandwidth for the mean-shift algorithm: the mean-shift algorithm is a gradient ascent algorithm, therefore it appears more natural to select the bandwidth in such a way to optimize the estimation of the density derivatives (\citealp{chacon2013data}). An incomplete list of other proposals includes \cite{comaniciuadaptivemeanshift}, who propose strategies to select data-driven adaptive bandwidths, \cite{einbeck2011bandwidth}, who introduces a bandwidth selector based on \textit{principal curves} (\citealp{hastie1989principal}; \citealp{flury1990principal}; \citealp{flury1993estimation}), and \cite{genovese2013nonparametric}, who propose a bandwidth selection method based on the maximization of the number of significant modes.

The significance test of Section \ref{sec:inference} allows us to choose the bandwidth in such a way to maximize the number of significant modes as suggested in \cite{genovese2013nonparametric}. Also, the multi-bandwidth analysis that we discuss in Sections \ref{sec:neural} and \ref{sec:signatures} offers an alternative approach for the selection of the bandwidth in applied work. However, it seems harder to come up with a theory for bandwidth selection in the infinite-dimensional setting that we consider. We regard the investigation of theoretically justified data-driven bandwidth selectors for the functional mean-shift algorithm as an interesting direction for further theoretical research.

\acks{ADD ACKNOWLEDGMENTS}

\newpage

\appendix

\section*{Proof of Lemmata}
\textbf{Proof of Lemma \ref{lemma:gateaux}}\\
The Gate\^aux differential of the estimated surrogate density can be obtained by applying the chain rule. We report here a more explicit calculation that highlights the relationschip between the kernel $K$ used in the functional mean-shift algorithm and its shadow $G$ that is used in the corresponding gradient ascent analog.\\
\begin{align*}
		\tilde p_x(y) 	&= \left. \frac{d}{d \alpha} \tilde p(x+\alpha y) \right|_{\alpha=0} = \left. \frac{d}{d \alpha} w_G(\mathcal{S}) \sum_{X \in \mathcal{S}} G_h \left( X,x+\alpha y)\right) \right|_{\alpha=0} \\
				&= \left. \frac{d}{d \alpha} w_G(\mathcal{S}) \sum_{X \in \mathcal{S}} g \left( \frac{d(X, x+\alpha y)}{h(X)}\right) \right|_{\alpha=0} \\
				&= \left. \frac{d}{d \alpha} w_G(\mathcal{S}) \sum_{X \in \mathcal{S}} g \left( \frac{\|X - x -\alpha y\|}{h(X)}\right) \right|_{\alpha=0} \\
				&= \left. w_G(\mathcal{S}) \sum_{X \in \mathcal{S}} - \frac{1}{h(X)}g' \left( \frac{\| X - x -\alpha y\|)}{h(X)}\right) \frac{\alpha \|y\|^2 - \langle X-x,y\rangle}{\|X-x-\alpha y\|}  \right|_{\alpha=0} \\
				&= \left \langle y, w_G(\mathcal{S}) \sum_{X \in \mathcal{S}} - \frac{1}{h(X)}g' \left( \frac{d(X, x)}{h(X)}\right) \frac{X-x}{d(X, x)} \right \rangle \\
				&= \left \langle y, Cw_G(\mathcal{S}) \sum_{X \in \mathcal{S}}\frac{1}{h^2(X)} k \left( \frac{d(X,x)}{h(X)}\right) \left(X-x\right) \right \rangle \\
				&= \left \langle y, Cw_G(\mathcal{S}) \sum_{X \in \mathcal{S}} \frac{1}{h^2(X)} K_h \left( X,x\right) \left(X-x\right) \right \rangle.
	\end{align*}

\newpage

\noindent \textbf{Proof of Lemma \ref{lemma:secondGateauxdifferential}}\\
%
\begin{align*}
	\tilde p^{(2)}_x(y,z) &= \left. \frac{d}{d\alpha} \tilde p_{x+\alpha z}(y) \right|_{\alpha=0} \\
	&=\left. \frac{d}{d\alpha} \left \langle y, C w_G(\mathcal{S}) \sum_{X \in \mathcal{S}} \frac{1}{h^2(X)} K_h\left(X,x+\alpha z \right) \left(X-x-\alpha z\right) \right \rangle \right|_{\alpha=0} \\
					&= \left. \left \langle y, C w_G(\mathcal{S}) \sum_{X \in \mathcal{S}} \frac{1}{h^2(X)} \frac{d}{d\alpha} K_h\left(X,x+\alpha z \right) \left(X-x-\alpha z\right) \right \rangle \right|_{\alpha=0} \\
					&= \left. \Bigg \langle y, C w_G(\mathcal{S}) \sum_{X \in \mathcal{S}} \frac{1}{h^2(X)} \Bigg[- K_h(X,x+\alpha z) z  \right.\\
					& - \left. \frac{1}{h(X)} \frac{\langle X-x,z \rangle + \alpha \|z\|^2}{\|X-x-\alpha z\|}K_h'(X,x+\alpha z)(X-x-\alpha z) \Bigg] \Bigg \rangle \right|_{\alpha=0} \\
					&= \Bigg \langle y, C w_G(\mathcal{S}) \sum_{X \in \mathcal{S}} \frac{1}{h^2(X)} \Bigg[ -\frac{1}{h(X)} \frac{\langle X-x,z \rangle}{\|X-x\|}K_h'(X,x)(X-x) \\
					&- K_h(X,x) z \Bigg] \Bigg \rangle\\
					&= - \Bigg \langle y, C w_G(\mathcal{S}) \sum_{X \in \mathcal{S}} \frac{1}{h^2(X)} \Bigg[\frac{1}{h(X)} \frac{\langle X-x,z \rangle}{\|X-x\|}K_h'(X,x)(X-x) \\
					& + K_h(X,x) z \Bigg] \Bigg \rangle\\
					&= - C w_G(\mathcal{S}) \sum_{X \in \mathcal{S}} \frac{1}{h^2(X)} \Bigg[\frac{1}{h(X)} K_h'(X,x) \frac{\langle X-x,y \rangle \langle X-x,z \rangle}{\|X-x\|} \\
					&+ K_h(X,x) \langle y, z \rangle \Bigg].
\end{align*}

\newpage

\noindent \textbf{Proof of Lemma \ref{lemma:test statistics}}\\
	\begin{align*}
		\tilde \lambda_x &= \sup_{\|y\|=1} \tilde p^{(2)}_x(y,y) = \\
				&= \sup_{\|y\|=1} - C w_G(\mathcal{S}) \sum_{X \in \mathcal{S}} \frac{1}{h^2(X)} \Bigg[\frac{1}{h(X)} K_h'(X,x) \frac{\langle X-x,y \rangle \langle X-x,y \rangle}{\|X-x\|} \\
				& + K_h(X,x) \langle y, y \rangle \Bigg] \\
				&= \sup_{\|y\|=1} - C w_G(\mathcal{S}) \sum_{X \in \mathcal{S}} \frac{1}{h^2(X)} \Bigg[\frac{1}{h(X)} K_h'(X,x) \frac{\|X-x\|^2 + \|y\|^2 + 2 \langle X-x,y\rangle}{\|X-x\|}\\
				& + K_h(X,x) \|y\|^2 \rangle \Bigg] \\
				&= - C w_G(\mathcal{S}) \sum_{X \in \mathcal{S}} \frac{1}{h^2(X)} \Bigg[\frac{1}{h(X)} K_h'(X,x) \left[ \|X-x\| + \|X-x\|^{-1} \right] + K_h(X,x) \rangle \Bigg] \\
				& + \sup_{\|y\|=1} -C w_G(\mathcal{S}) \sum_{X \in \mathcal{S}} \frac{2}{h^3(X)}K'_h(X,x) \frac{\langle X-x,y\rangle}{\|X-x\|} \\
				&=  - C w_G(\mathcal{S}) \sum_{X \in \mathcal{S}} \frac{1}{h^2(X)} \Bigg[\frac{1}{h(X)} K_h'(X,x) \left[ \|X-x\| + \|X-x\|^{-1} \right] \\
				& + K_h(X,x) \rangle \Bigg] +  \sup_{\|y\|=1} \left \langle y, -C w_G(\mathcal{S}) \sum_{X \in \mathcal{S}} \frac{2}{h^3(X)}K'_h(X,x) \frac{X-x}{\|X-x\|} \right \rangle \\
				&=   - C w_G(\mathcal{S}) \sum_{X \in \mathcal{S}} \frac{1}{h^2(X)} \Bigg[\frac{1}{h(X)} K_h'(X,x) \left[ \|X-x\| + \|X-x\|^{-1} \right] \\
				& + K_h(X,x) \rangle \Bigg] + 2 C w_G(\mathcal{S}) \left \| \sum_{X \in \mathcal{S}} \frac{1}{h^3(X)} K'_h(X,x) \frac{X-x}{\|X-x\|} \right \| \\
				&= Cw_G(\mathcal{S}) \left[ 2 \left\| \sum_{X \in \mathcal{S}} \frac{1}{h^3(X)}K'_h(X,x)\frac{X-x}{\|X-x\|}\right\| -  \right.\\\
				&  \left. \sum_{X \in \mathcal{S}} \frac{1}{h^2(X)}\left[ \frac{1}{h(X)}K'_h(X,x)\left( \|X-x\| + \|X-x\|^{-1} \right) + K_h(X,x)\right]\right].
	\end{align*}

\newpage

\bibliography{MyBibliography.bib}

\begin{thebibliography}{51}
\providecommand{\natexlab}[1]{#1}
\providecommand{\url}[1]{\texttt{#1}}
\expandafter\ifx\csname urlstyle\endcsname\relax
  \providecommand{\doi}[1]{doi: #1}\else
  \providecommand{\doi}{doi: \begingroup \urlstyle{rm}\Url}\fi

\bibitem[Aliyari~Ghassabeh(2013)]{AliyariGhassabehOnConvergence}
Youness Aliyari~Ghassabeh.
\newblock On the convergence of the mean shift algorithm in the one-dimensional
  space.
\newblock \emph{Pattern Recognition Letters}, 34\penalty0 (12):\penalty0 1423
  -- 1427, 2013.

\bibitem[Ambrosetti and Prodi(1995)]{ambrosetti1995primer}
Antonio Ambrosetti and Giovanni Prodi.
\newblock \emph{A primer of nonlinear analysis}.
\newblock Number~34. Cambridge University Press, 1995.

\bibitem[Arias-Castro et~al.(2013)Arias-Castro, Mason, and
  Pelletier]{ariascastromeanshiftgradient}
Ery Arias-Castro, David Mason, and Bruno Pelletier.
\newblock On the estimation of the gradient lines of a density and the
  consistency of the mean-shift algorithm.
\newblock \emph{unpublished manuscript}, 2013.

\bibitem[Azzalini and Torelli(2007)]{azzalini2007clustering}
Adelchi Azzalini and Nicola Torelli.
\newblock Clustering via nonparametric density estimation.
\newblock \emph{Statistics and Computing}, 17\penalty0 (1):\penalty0 71--80,
  2007.

\bibitem[Bosq(2000)]{bosq2000linear}
Denis Bosq.
\newblock \emph{Linear processes in function spaces: theory and applications}.
\newblock Springer, 2000.

\bibitem[Carreira-Perpi{\~n}{\'a}n(2006)]{carreira2006fast}
Miguel~{\'A}. Carreira-Perpi{\~n}{\'a}n.
\newblock Fast nonparametric clustering with gaussian blurring mean-shift.
\newblock In \emph{Proceedings of the 23rd International Conference on Machine
  Learning}, pages 153--160, 2006.

\bibitem[Chac{\'o}n(2012)]{chacon2012clusters}
Jos{\'e}~E. Chac{\'o}n.
\newblock Clusters and water flows: a novel approach to modal clustering
  through morse theory.
\newblock \emph{arXiv preprint arXiv:1212.1384}, 2012.

\bibitem[Chac{\'o}n(2014)]{chacon2012populationbackground}
Jos{\'e}~E. Chac{\'o}n.
\newblock A population background for nonparametric density-based clustering.
\newblock \emph{unpublished manuscript}, 2014.

\bibitem[Chac{\'o}n and Duong(2013)]{chacon2013data}
Jos{\'e}~E. Chac{\'o}n and Tarn Duong.
\newblock Data-driven density derivative estimation, with applications to
  nonparametric clustering and bump hunting.
\newblock \emph{Electronic Journal of Statistics}, 7:\penalty0 499--532, 2013.

\bibitem[Chazal et~al.(2013)Chazal, Guibas, Oudot, and
  Skraba]{chazal2013persistence}
Fr{\'e}d{\'e}ric Chazal, Leonidas~J Guibas, Steve~Y Oudot, and Primoz Skraba.
\newblock Persistence-based clustering in riemannian manifolds.
\newblock \emph{Journal of the ACM (JACM)}, 60\penalty0 (6):\penalty0 41, 2013.

\bibitem[Cheng(1995)]{chengmeanshift}
Yizong Cheng.
\newblock Mean shift, mode seeking, and clustering.
\newblock \emph{IEEE Transactions on Pattern Analysis and Machine
  Intelligence}, 17\penalty0 (8):\penalty0 790--799, 1995.

\bibitem[Chernoff(1964)]{chernoff1964mode}
Herman Chernoff.
\newblock Estimation of the mode.
\newblock \emph{Annals of the Institute of Statistical Mathematics},
  16\penalty0 (1):\penalty0 31--41, 1964.

\bibitem[Comaniciu and Meer(2002)]{comaniciu02meanshift}
Dorin Comaniciu and Peter Meer.
\newblock Mean shift: a robust approach toward feature space analysis.
\newblock \emph{IEEE Transactions on Pattern Analysis and Machine
  Intelligence}, 24:\penalty0 603--619, 2002.

\bibitem[Comaniciu et~al.(2001)Comaniciu, Ramesh, and
  Meer]{comaniciuadaptivemeanshift}
Dorin Comaniciu, Visvanathan Ramesh, and Peter Meer.
\newblock The variable bandwidth mean shift and data-driven scale selection.
\newblock In \emph{Proceedings of the Eighth IEEE International Conference on
  Computer Vision}, volume~1, pages 438--445, 2001.

\bibitem[Cuevas and Fraiman(1997)]{cuevas1997plug}
Antonio Cuevas and Ricardo Fraiman.
\newblock A plug-in approach to support estimation.
\newblock \emph{The Annals of Statistics}, 25\penalty0 (6):\penalty0
  2300--2312, 1997.

\bibitem[Dabo-Niang et~al.(2004)Dabo-Niang, Ferraty, and
  Vieu]{dabo2004estimation}
Sophie Dabo-Niang, Fr{\'e}d{\'e}ric Ferraty, and Philippe Vieu.
\newblock Estimation du mode dans un espace vectoriel semi-norm{\'e}.
\newblock \emph{Comptes Rendus Mathematique}, 339\penalty0 (9):\penalty0
  659--662, 2004.

\bibitem[Delaigle and Hall(2010)]{delaigle2010}
Aurore Delaigle and Peter Hall.
\newblock Defining probability density for a distribution of random functions.
\newblock \emph{The Annals of Statistics}, 38\penalty0 (2):\penalty0
  1171--1193, 2010.

\bibitem[Donoho and Liu(1991)]{donoho1991geometrizing}
David~L. Donoho and Richard~C. Liu.
\newblock Geometrizing rates of convergence, iii.
\newblock \emph{The Annals of Statistics}, pages 668--701, 1991.

\bibitem[Eddy(1980)]{eddy1980}
William~F. Eddy.
\newblock Optimum kernel estimators of the mode.
\newblock \emph{The Annals of Statistics}, 8\penalty0 (4):\penalty0 870--882,
  1980.

\bibitem[Einbeck(2011)]{einbeck2011bandwidth}
Jochen Einbeck.
\newblock Bandwidth selection for mean-shift based unsupervised learning
  techniques: a unified approach via self-coverage.
\newblock \emph{Journal of Pattern Recognition Research}, 6\penalty0
  (2):\penalty0 175--192, 2011.

\bibitem[Faraway and Jhun(1990)]{faraway1990bootstrap}
Julian~J. Faraway and Myoungshic Jhun.
\newblock Bootstrap choice of bandwidth for density estimation.
\newblock \emph{Journal of the American Statistical Association}, 85\penalty0
  (412):\penalty0 1119--1122, 1990.

\bibitem[Ferraty and Vieu(2006)]{ferraty2006nonparametric}
Fr\'ed\'eric Ferraty and Philippe Vieu.
\newblock \emph{Nonparametric functional data analysis: theory and practice}.
\newblock Springer, 2006.

\bibitem[Ferraty et~al.(2006)Ferraty, Laksaci, and Vieu]{estimating2006ferraty}
Fr\'ed\'eric Ferraty, Ali Laksaci, and Philippe Vieu.
\newblock Estimating some characteristics of the conditional distribution in
  nonparametric functional models.
\newblock \emph{Statistical Inference for Stochastic Processes}, 9\penalty0
  (1):\penalty0 47--76, 2006.

\bibitem[Ferraty et~al.(2012)Ferraty, Kudraszow, and
  Vieu]{ferratysurrogatedensity}
Fr\'ed\'eric Ferraty, Nadia Kudraszow, and Philippe Vieu.
\newblock Nonparametric estimation of a surrogate density function in
  infinite-dimensional spaces.
\newblock \emph{Journal of Nonparametric Statistics}, 24\penalty0 (2):\penalty0
  447--464, 2012.

\bibitem[Flury(1993)]{flury1993estimation}
Bernard~D. Flury.
\newblock Estimation of principal points.
\newblock \emph{Applied Statistics}, pages 139--151, 1993.

\bibitem[Flury(1990)]{flury1990principal}
Bernhard~A. Flury.
\newblock Principal points.
\newblock \emph{Biometrika}, 77\penalty0 (1):\penalty0 33--41, 1990.

\bibitem[Fukunaga and Hostetler(1975)]{fukunagahostetlermeanshift}
Keinosuke Fukunaga and Larry Hostetler.
\newblock The estimation of the gradient of a density function, with
  applications in pattern recognition.
\newblock \emph{IEEE Transactions on Information Theory}, 21\penalty0
  (1):\penalty0 32--40, 1975.

\bibitem[Gasser et~al.(1998)Gasser, Hall, and
  Presnell]{gasser1998nonparametric}
Theo Gasser, Peter Hall, and Brett Presnell.
\newblock Nonparametric estimation of the mode of a distribution of random
  curves.
\newblock \emph{Journal of the Royal Statistical Society, Series B},
  60\penalty0 (4):\penalty0 681--691, 1998.

\bibitem[Geenens(2011)]{geenenssignatures}
Gery Geenens.
\newblock A nonparametric functional method for signature recognition.
\newblock In Fr\'ed\'eric Ferraty, editor, \emph{Recent advances in functional
  data analysis and related topics}, Contributions to statistics, pages
  141--147. Physica-Verlag HD, 2011.

\bibitem[Genovese et~al.(2013)Genovese, Perone-Pacifico, Verdinelli, and
  Wasserman]{genovese2013nonparametric}
Christopher Genovese, Marco Perone-Pacifico, Isabella Verdinelli, and Larry
  Wasserman.
\newblock Nonparametric inference for density modes.
\newblock \emph{arXiv preprint arXiv:1312.7567}, 2013.

\bibitem[Hartigan(1975)]{hartigan1975clustering}
John~A. Hartigan.
\newblock \emph{Clustering algorithms}.
\newblock John Wiley \& Sons, 1975.

\bibitem[Hastie and Stuetzle(1989)]{hastie1989principal}
Trevor Hastie and Werner Stuetzle.
\newblock Principal curves.
\newblock \emph{Journal of the American Statistical Association}, 84\penalty0
  (406):\penalty0 502--516, 1989.

\bibitem[Horv{\'a}th and Kokoszka(2012)]{horvath2012inference}
Lajos Horv{\'a}th and Piotr Kokoszka.
\newblock \emph{Inference for functional data with applications}.
\newblock Springer, 2012.

\bibitem[Jacques and Preda(2013)]{jacques2013functional}
Julien Jacques and Cristian Preda.
\newblock Functional data clustering: a survey.
\newblock \emph{Advances in Data Analysis and Classification}, pages 1--25,
  2013.

\bibitem[Klemel{\"a}(2005)]{klemela2005adaptive}
Jussi Klemel{\"a}.
\newblock Adaptive estimation of the mode of a multivariate density.
\newblock \emph{Journal of Nonparametric Statistics}, 17\penalty0 (1):\penalty0
  83--105, 2005.

\bibitem[Lei et~al.(2013)Lei, Rinaldo, and Wasserman]{conformalexplore2013lei}
Jing Lei, Alessandro Rinaldo, and Larry Wasserman.
\newblock A conformal prediction approach to explore functional data.
\newblock \emph{Annals of Mathematics and Artificial Intelligence}, pages
  1--15, 2013.

\bibitem[Li et~al.(2007{\natexlab{a}})Li, Ray, and
  Lindsay]{li2007nonparametric}
Jia Li, Surajit Ray, and Bruce~G. Lindsay.
\newblock A nonparametric statistical approach to clustering via mode
  identification.
\newblock \emph{Journal of Machine Learning Research}, 8\penalty0 (8):\penalty0
  1687--1723, 2007{\natexlab{a}}.

\bibitem[Li et~al.(2007{\natexlab{b}})Li, Hu, and Wu]{LiNoteOnConvergence}
Xiangru Li, Zhanyi Hu, and Fuchao Wu.
\newblock A note on the convergence of the mean shift.
\newblock \emph{Pattern Recognition}, 40\penalty0 (6):\penalty0 1756 -- 1762,
  2007{\natexlab{b}}.

\bibitem[Park and Marron(1990)]{park1990comparison}
Byeong~U. Park and James~S. Marron.
\newblock Comparison of data-driven bandwidth selectors.
\newblock \emph{Journal of the American Statistical Association}, 85\penalty0
  (409):\penalty0 66--72, 1990.

\bibitem[Parzen(1962)]{parzen1962estimation}
Emanuel Parzen.
\newblock On estimation of a probability density function and mode.
\newblock \emph{The Annals of Mathematical Statistics}, 33\penalty0
  (3):\penalty0 1065--1076, 1962.

\bibitem[Ramsay and Silverman(2005)]{ramsay2006functional}
James~O. Ramsay and Bernard~W. Silverman.
\newblock \emph{Functional Data Analysis}.
\newblock Springer, 2005.

\bibitem[Rinaldo and Wasserman(2010)]{rinaldo2010generalized}
Alessandro Rinaldo and Larry Wasserman.
\newblock Generalized density clustering.
\newblock \emph{The Annals of Statistics}, 38\penalty0 (5):\penalty0
  2678--2722, 2010.

\bibitem[Rinaldo et~al.(2012)Rinaldo, Singh, Nugent, and
  Wasserman]{rinaldo2012stability}
Alessandro Rinaldo, Aarti Singh, Rebecca Nugent, and Larry Wasserman.
\newblock Stability of density-based clustering.
\newblock \emph{Journal of Machine Learning Research}, 13\penalty0
  (1):\penalty0 905--948, 2012.

\bibitem[Romano(1988)]{romano1988}
Joseph~P. Romano.
\newblock On weak convergence and optimality of kernel density estimates of the
  mode.
\newblock \emph{The Annals of Statistics}, 16\penalty0 (2):\penalty0 629--647,
  1988.

\bibitem[Sheather and Jones(1991)]{sheather1991reliable}
Simon~J. Sheather and Michael~C. Jones.
\newblock A reliable data-based bandwidth selection method for kernel density
  estimation.
\newblock \emph{Journal of the Royal Statistical Society, Series B},
  53\penalty0 (3):\penalty0 683--690, 1991.

\bibitem[Silverman(1981)]{silverman1981using}
Bernard~W. Silverman.
\newblock Using kernel density estimates to investigate multimodality.
\newblock \emph{Journal of the Royal Statistical Society, Series B}, pages
  97--99, 1981.

\bibitem[Silverman(1986)]{Silverman86}
Bernard~W. Silverman.
\newblock \emph{Density estimation for statistics and data analysis}.
\newblock Chapman \& Hall, 1986.

\bibitem[Stuetzle(2003)]{stuetzle2003estimating}
Werner Stuetzle.
\newblock Estimating the cluster tree of a density by analyzing the minimal
  spanning tree of a sample.
\newblock \emph{Journal of Classification}, 20\penalty0 (1):\penalty0 025--047,
  2003.

\bibitem[Stuetzle and Nugent(2010)]{stuetzle2010generalized}
Werner Stuetzle and Rebecca Nugent.
\newblock A generalized single linkage method for estimating the cluster tree
  of a density.
\newblock \emph{Journal of Computational and Graphical Statistics}, 19\penalty0
  (2), 2010.

\bibitem[Vieu(1996)]{vieu1996note}
Philippe Vieu.
\newblock A note on density mode estimation.
\newblock \emph{Statistics \& Probability Letters}, 26\penalty0 (4):\penalty0
  297--307, 1996.

\bibitem[Vovk et~al.(2009)Vovk, Nouretdinov, and Gammerman]{vovk2009line}
Vladimir Vovk, Ilia Nouretdinov, and Alex Gammerman.
\newblock On-line predictive linear regression.
\newblock \emph{The Annals of Statistics}, 37\penalty0 (3):\penalty0
  1566--1590, 2009.

\end{thebibliography}

\end{document}